\documentclass[twocolumn]{autart}
\usepackage{mathtools}
\usepackage{amssymb}
\usepackage{balance}
\usepackage{enumitem}
\usepackage{mathdots}
\usepackage{xfrac}
\usepackage{bm}
\usepackage{bbm}
\usepackage{array}
\usepackage{algorithm}
\usepackage{algorithmicx}
\usepackage{algpseudocode}
\usepackage{comment}
\usepackage{pgfplots}
\usepgflibrary{shapes}
\usepackage{tikz}

\usetikzlibrary{arrows,calc,patterns}
\tikzstyle{Rect}=[draw=gray,line width=0.001pt,preaction={clip, postaction={pattern=north east lines, pattern color=gray,line width=0.1pt}}]
\tikzset{
	>=stealth',
	help lines/.style={dashed, thick},
	axis/.style={<->},
	important line/.style={thick},
	connection/.style={thick, dotted},
}
\newcommand{\tikzline}[1]{(\protect\tikz[baseline=-0.6ex,x=1pt,y=1pt]{ \protect\draw[#1] [-] (0,0) -- (10,0);})}
\newcommand{\tikzdashedline}[1]{(\protect\tikz[baseline=-0.6ex,x=0.9pt,y=1pt]{ \protect\draw[#1,dashed] [-] (0,0) -- (10,0);})}

\definecolor{MatlabBlue}{rgb}    {0     , 0.4470, 0.7410}
\definecolor{MatlabRed}{rgb}     {0.8500, 0.3250, 0.0980}
\definecolor{MatlabYellow}{rgb}  {0.9290, 0.6940, 0.1250}
\definecolor{MatlabPurple}{rgb}  {0.4940, 0.1840, 0.5560}
\definecolor{MatlabGreen}{rgb}   {0.4660, 0.6740, 0.1880}
\definecolor{MatlabBabyBlue}{rgb}{0.3010, 0.7450, 0.9330}
\definecolor{MatlabGray}{rgb}{0.5, 0.5, 0.5}
\definecolor{MatlabLightGray}{rgb}{0.75, 0.75, 0.75}
\definecolor{MatlabBlack}{rgb}{0, 0, 0}
\definecolor{MatlabLightGray4}{rgb}{0.875, 0.875, 0.875}
\definecolor{MatlabLightGray3}{rgb}{0.85, 0.85, 0.85}
\definecolor{MatlabLightGray2}{rgb}{0.775, 0.775, 0.775}
\definecolor{MatlabLightGray1}{rgb}{0.7, 0.7, 0.7}
\definecolor{MatlabGray20}{rgb}{0.2, 0.2, 0.2}
\definecolor{MatlabGray30}{rgb}{0.3, 0.3, 0.3}
\definecolor{MatlabGray40}{rgb}{0.4, 0.4, 0.4}
\definecolor{MatlabGray50}{rgb}{0.5, 0.5, 0.5}
\definecolor{MatlabGray60}{rgb}{0.6, 0.6, 0.6}
\definecolor{MatlabGray70}{rgb}{0.7, 0.7, 0.7}
\definecolor{MatlabGray80}{rgb}{0.8, 0.8, 0.8}
\definecolor{MatlabGray85}{rgb}{0.85, 0.85, 0.85}
\definecolor{MatlabGray90}{rgb}{0.9, 0.9, 0.9}

\newtheorem{assumption}{Assumption}
\newtheorem{lemma}{Lemma}
\newtheorem{corollary}{Corollary}

\newtheorem{theorem}{Theorem}

\usepackage{etoolbox}

\makeatletter
\def\@xnamedef#1{\expandafter\protected@xdef\csname #1\endcsname}
\def\no@harm{} 
\def\ead@au#1{\protected@edef\@ead@au{#1}}
\patchcmd\runningauthor@fmt{\global\edef}{\protected@xdef}{}{}
\patchcmd\runningauthor@fmt{\global\edef}{\protected@xdef}{}{}
\patchcmd\author@fmt{\edef}{\protected@edef}{}{}
\patchcmd\add@xtok{\xdef}{\protected@xdef}{}{}
\makeatother
                                           
\usepackage{color}
\usepackage{eso-pic}
\AddToShipoutPictureBG*{%
	\AtPageUpperLeft{%
		\setlength\unitlength{1in}%
		\makebox(8,-0.75)[c]{
			\begin{tabular}{c c}
				Rodrigo A. Gonz\'alez \emph{et al.},
				Identification of Additive Continuous-time Systems in Open and Closed Loop \\
				To appear in
				{\em Automatica},
				uploaded to Arxiv on September 25th, 2024 \\
\end{tabular}}}}

\begin{document}
	\begin{frontmatter}
		\title{Identification of Additive Continuous-time Systems in Open and Closed Loop\thanksref{footnoteinfo}} 
		
		\thanks[footnoteinfo]{The material in this paper was not presented at any IFAC meeting. Corresponding author: Rodrigo A. Gonz\'alez.}
		
		\author[TUE]{Rodrigo A. Gonz\'alez}\ead{r.a.gonzalez@tue.nl},    
		\author[TUE]{Koen Classens}\ead{k.h.j.classens@tue.nl}, %
            \author[KTH]{Cristian R. Rojas}\ead{crro@kth.se},    
		\author[UON]{James S. Welsh}\ead{james.welsh@newcastle.edu.au},
		\author[TUE,Delft]{Tom Oomen}\ead{t.a.e.oomen@tue.nl}   
		\address[TUE]{Control Systems Technology Section, Department of Mechanical Engineering, Eindhoven University of Technology, Eindhoven, The Netherlands.}  
		\address[KTH]{Division of Decision and Control Systems, KTH Royal Institute of Technology, 10044 Stockholm, Sweden.}  
		\address[UON]{School of Engineering, University of Newcastle, Callaghan, 2308 NSW, Australia.} 
		\address[Delft]{Delft Center for Systems and Control, Delft University of Technology, Delft, The Netherlands.}  
		
		
		\begin{abstract}
When identifying electrical, mechanical, or biological systems, parametric continuous-time identification methods can lead to interpretable and parsimonious models when the model structure aligns with the physical properties of the system. Traditional linear system identification may not consider the most parsimonious model when relying solely on unfactored transfer functions, which typically result from standard direct approaches. This paper presents a novel identification method that delivers additive models for both open and closed-loop setups. The estimators that are derived are shown to be generically consistent, and can admit the identification of marginally stable additive systems. Numerical simulations show the efficacy of the proposed approach, and its performance in identifying a modal representation of a flexible beam is verified using experimental data.	
		\end{abstract}
		\end{frontmatter}
	
	\section{Introduction}
The goal in system identification is to obtain mathematical descriptions of systems using input and output data. These models can play a critical role for prediction, analysis, and design of control laws for electrical, mechanical, biological, or environmental systems, among others \cite{soderstrom1989system,ljung1998system}. A distinction can be made between discrete-time and continuous-time system identification methods using sampled data. Identification algorithms for discrete-time systems provide difference equation models that give meaningful information only at the sampling instants. In contrast, continuous-time system identification methods use sampled data to estimate differential equation models that represent the system at any point in time.
 
 Linear continuous-time system identification methods \cite{rao2006identification} are widely used and successful in a range of practical applications \cite{young2012recursive,garnier2014advantages,garnier2014direct}, offering several advantages over the standard discrete-time algorithms. One advantage of continuous-time methods is their ability to directly incorporate \textit{a priori} knowledge of the relative degree of the physical systems they model. This feature is particularly useful in estimating, e.g., mechanical systems, as they often exhibit no impulse response discontinuities due to the double integration relationship between force and position \cite{gawronski2004advanced}, leading to models with relative degree equal to 2. In contrast, discrete-time methods must also account for sampling zeros that are not present in the continuous-time system representation \cite{aastrom1984zeros}, requiring an additional optimization step if the resulting discrete-time model is later converted to continuous time \cite{gonzalez2018asymptotically}. Recent research has proposed tailored variants of instrumental variables to address this particular challenge~\cite{gonzalez2023relation}.

When addressing the identification of physical systems, increasingly stringent performance requirements necessitate the use of parsimonious models, i.e., models with the fewest number of parameters that can adequately describe the phenomenon under investigation. From a statistical standpoint, it is well known that parsimonious model structures that contain the true system within the model set result in reduced variance compared to over-parameterized model sets \cite{stoica1982parsimony}. Most linear continuous-time identification methods parameterize the model structure as an unfactored transfer function, that is, a quotient of polynomials in the Laplace transform variable $s$ with numerator and denominator polynomial degrees indicating the number of zeros and poles of the model, respectively. Unfactored transfer function parametrizations may not always offer the most parsimonious model structure, but due to their simplicity they are widely adopted in the Simplified Refined Instrumental Variable method for Continuous-time systems (SRIVC, \cite{young1980refined}), the Poisson moment functional approach \cite{saha1982general}, the Least-Squares State-Variable Filter method (LSSVF, \cite{young1965determination}), and other linear filter and integral methods \cite{garnier2003continuous}. The closed-loop variants of these estimators are also limited to estimating unfactored transfer functions \cite{gilson2003continuous,gilson2008instrumental,young2009simple}. Although more general model structures can be incorporated using the prediction error method and maximum likelihood paradigm \cite{astrom1979maximum}, typically one unfactored transfer function is estimated in the linear and time-invariant continuous-time identification problem~\cite{ljung2009experiments}.

While unfactored transfer functions are standard choices for the model structure in linear system identification, many practical applications related to, e.g., flexible motion systems \cite{oomen2018advanced} and vibration analysis \cite{gawronski2004advanced}, often involve systems that are more easily interpreted as a sum of transfer functions with distinct denominators, typically corresponding to different resonant modes. In addition, modal parameter estimation for structures subject to vibrations is essential for design and model validation, to guarrantee safety and quality control \cite{voorhoeve2020identifying,reynders2012system}. Additive model parametrizations have advantages such as leading to physically more insightful models for fault diagnosis \cite{classens2022fault,classens2023fault} and improving the numerical conditioning of parameter estimation for high-order or highly-resonant systems \cite{gilson2017frequency}. These model parametrizations, which have previously been considered in statistics \cite{hastie1986generalized} and econometrics \cite{hardle2004bootstrap}, offer increased model flexibility and the ability to decentralize the analysis of each additive component for optimization and control purposes. Despite some contributions in nonlinear discrete-time finite-impulse response and generalized Hammerstein model estimation \cite{bai2005identification,bai2008identification}, the application of additive model structures in the realm of system identification has been limited. Recently, \cite{gonzalez2023parsimonious} introduced a block coordinate algorithm to identify continuous-time systems under an additive model structure. However, this approach is confined to open-loop setups and can be computationally demanding due to the need for estimating each submodel with a tailored version of the SRIVC method at each iteration of the descent algorithm \cite{gonzalez2024statistical}.

Another difficulty when estimating linear systems with an additive model decomposition is that some systems are known to have integral action, i.e., their transfer function descriptions contain integrators. Applications featuring such systems can be found in platooning \cite{ploeg2011design}, hydraulics, mechanical systems \cite{gawronski2004advanced,preumont2018vibration}, among others. As an illustrative example, mechatronic positioning systems can often be represented as a combination of rigid-body and flexible modes, with the rigid-body modes modeled as double integrators \cite{oomen2018advanced}. Many system identification methods are only suitable for asymptotically stable systems, since the predictors become ill-conditioned when the models are unstable. Although this problem has been tackled for Box-Jenkins model structures in discrete-time and continuous-time settings \cite{forssell2000identification,gonzalez2022unstable}, these works do not consider additive model parametrizations.

In this paper, we propose a comprehensive identification method for modeling additive linear continuous-time systems in both open and closed-loop settings. Our contributions can be summarized as follows:
\begin{enumerate}[label=C\arabic*]
\label{contribution1}
	\item 
    We derive the optimality conditions that the proposed estimators for additive continuous-time system identification must satisfy in both open and closed-loop scenarios in a unified manner. To achieve this, we establish a connection between the first-order optimality condition of the open-loop estimator in an output error setup and the instrumental variable approach in the closed-loop setting. In the closed-loop case, we derive explicit expressions for an instrument vector that ensures a consistent estimator while also yielding a minimum asymptotic covariance matrix in a positive definite sense. The closed-loop results extend the ones found in \cite{gilson2008instrumental} to a general class of additive continuous-time models.
\label{contribution2}
    \item 
    We develop open and closed-loop estimators based on the derived optimality conditions, extending the SRIVC and CLSRIVC estimators for additive continuous-time models. We also consider the identification of marginally stable additive systems, and provide a thorough consistency analysis for our estimators, demonstrating their generic consistency under mild conditions.
	
	\item We evaluate the proposed method through extensive Monte Carlo simulations, and we show its efficacy using data from an experimental flexible beam setup.
\end{enumerate}

The remainder of this paper is structured as follows. Section \ref{sec:system} introduces the problem setup for both open and closed-loop settings. Section \ref{sec:optimality} describes the optimality conditions for the additive model structures of both settings. Section \ref{sec:parsimonious} contains the proposed unified iterative procedure for additive continuous-time system identification with its asymptotic analysis. Extensive simulations can be found in Section \ref{sec:simulations}, while Section \ref{sec:experiments} contains experimental setup results. Concluding remarks are presented in Section \ref{sec:conclusions}. A technical lemma used for proving the main theoretical results can be found in the Appendix.

\section{System and Model Setup}
\label{sec:system}
\begin{figure}
	\setlength{\unitlength}{0.092in} 
	\centering 
	\begin{picture}(37,21.5) 
		\put(18,19.5){\small{(a)}}
		\put(5.5,13.6){\vector(1,0){3.6}}
		\put(5.7,14.2){\footnotesize{$u(t_k)$}}
		\put(9.1,11.8){\framebox(3.6,3.6){\footnotesize{ZOH}}}
		\put(12.7,13.6){\vector(1,0){3.4}}
		\put(13.2,14.2){\footnotesize{$u(t)$}}
		\put(16.1,11.8){\framebox(4,3.6){\footnotesize{$G^*\hspace{-0.02cm}(p)$}}}
		\put(20.1,13.6){\vector(1,0){2.5}}
		\put(21.7,14.1){\scriptsize{+}}
		\put(22.5,18.3){\footnotesize{$v(t)$}}
		\put(23.5,13.6){\circle{1.6}}
		\put(23.5,17.8){\vector(0,-1){3.3}}
		\put(23.8,14.9) {\scriptsize{+}}
		\put(24.3,13.6){\line(1,0){0.8}}
		\put(25.1,13.6){\line(2,1){2}}
		\put(25.7,12.5){\footnotesize{$t_k$}}	
		\put(27.3,13.6){\vector(1,0){3.9}}
		\put(27.7,14.2){\footnotesize{$y(t_k)$}}	

		\put(0,4.1){\vector(1,0){4}}
		\put(0.1,4.7){\footnotesize{$r(t_k)$}}
		\put(3,4.6){\scriptsize{+}}
		\put(18,9.7){\small{(b)}}
		\put(3.7,2.7){\scriptsize{$-$}}
		\put(4.8,0){\vector(0,1){3.3}}
		\put(4.8,0){\line(1,0){29.7}}
		\put(4.8,4.1){\circle{1.6}}
		\put(5.6,4.1){\vector(1,0){1.8}}
		\put(7.4,2.3){\framebox(4,3.6){\small{$C_\textnormal{d}(q)$}}}
		\put(11.5,4.1){\vector(1,0){3.6}}
		\put(11.7,4.7){\footnotesize{$u(t_k)$}}
		\put(15.1,2.3){\framebox(3.6,3.6){\footnotesize{ZOH}}}
		\put(18.7,4.1){\vector(1,0){3.4}}
		\put(19.2,4.7){\footnotesize{$u(t)$}}
		\put(22.1,2.3){\framebox(4,3.6){\footnotesize{$G^*\hspace{-0.02cm}(p)$}}}
		\put(26.1,4.1){\vector(1,0){2}}
		\put(27.2,4.6){\scriptsize{+}}
		\put(28,8.8){\footnotesize{$v(t)$}}
		\put(29,4.1){\circle{1.6}}
		\put(29,8.3){\vector(0,-1){3.3}}
		\put(29.3,5.4) {\scriptsize{+}}
		\put(29.8,4.1){\line(1,0){0.8}}
		\put(30.6,4.1){\line(2,1){2}}
		\put(31.2,3){\footnotesize{$t_k$}}	
		\put(34.5,0){\line(0,1){4.1}}
		\put(32.8,4.1){\vector(1,0){3.9}}
		\put(33.2,4.7){\footnotesize{$y(t_k)$}}	
	\end{picture}
	\vspace{0.1cm}
	\caption{Block diagrams for the open (a) and closed-loop (b) settings studied in this paper.}
	\label{fig_setups}
\end{figure}
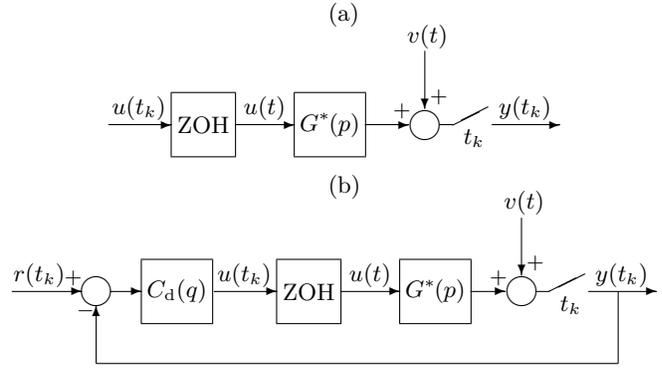

Consider the single-input single-output, linear and time-invariant (LTI), continuous-time system in additive form
\begin{equation}
	\label{ctsystem1}
	x(t) =  \sum_{i=1}^K G_i^*(p) u(t),
\end{equation}
where $p$ is the Heaviside or derivative operator (i.e., $pu(t)=\frac{\textnormal{d}}{\textnormal{d}t}u(t)$) and $u(t)$ is the input signal. Each subsystem $G_i^*(p)$ has $n_i$ poles and $m_i$ zeros, and can be expressed as $B_i^*(p)/A_i^*(p)$, where the numerator and denominator polynomials are assumed coprime, i.e., they do not share roots. These polynomials are given by
\begin{equation}
	\label{parametrization}
\begin{split}
	A_i^*(p)\hspace{-0.05cm}&= \hspace{-0.03cm}a_{i,n_i}^*p^{n_i}\hspace{-0.05cm}+\hspace{-0.05cm}a_{i,n_i-1}^*p^{n_i-1}\hspace{-0.05cm}+\cdots + a_{i,1}^*p +\hspace{-0.03cm} 1,  \hspace{-0.1cm}\\
	B_i^*(p)\hspace{-0.05cm}&=\hspace{-0.03cm} b_{i,m_i}^*p^{m_i}\hspace{-0.05cm}+\hspace{-0.05cm}b_{i,m+i-1}^*p^{m_i-1}\hspace{-0.05cm}+\cdots + b_{i,1}^*p \hspace{-0.03cm}+ \hspace{-0.03cm}b_{i,0}^*,	\hspace{-0.1cm}
\end{split}
\end{equation}
with $a_{n_i}^*\neq 0$ and $m_i\leq n_i$. We assume, without loss of generality, that the $A_i^*(p)$ polynomials do not share roots and that they are anti-monic, i.e., their constant coefficient is fixed to 1. In addition, to obtain a unique characterization of $\{G_i^*(p)\}_{i=1}^K$, we assume that at most one subsystem $G_i^*(p)$ has the same number of poles and zeros. The polynomials $A_i^*(p)$ and $B_i^*(p)$ are jointly described by the parameter vector 
\begin{equation}
	\label{ctparametervector}
	\bm{\theta}_i^* = \begin{bmatrix}
		a_{i,1}^*, &\hspace{-0.05cm} a_{i,2}^*, & \hspace{-0.05cm}\dots, & \hspace{-0.05cm}a_{i,n_i}^*, & \hspace{-0.05cm}b_{i,0}^*, & \hspace{-0.05cm}b_{i,1}^*, & \hspace{-0.05cm}\dots, & \hspace{-0.05cm}b_{i,m_i}^*
	\end{bmatrix}^\top.
\end{equation}
A noisy measurement of the output is retrieved at every time instant $t=t_k$, $k=1,\dots,N$, where $\{t_k\}_{k=1}^N$ are evenly spaced in time\footnote{Extensions of our proposed identification methods for irregularly sampled signals are feasible by following similar steps to the ones found in \cite{huselstein2002approach}.}. That is,
\begin{equation}
	\label{output}
	y(t_k) = x(t_k)+v(t_k),
\end{equation}
where $v(t_k)$ is assumed to be a zero-mean stationary random process of finite variance $\sigma^2$. Two frameworks are considered in this work: open and closed-loop identification. Block diagrams that describe these setups are presented in Figure~\ref{fig_setups}. The output noise $v(t_k)$ is assumed to be uncorrelated with the input samples $u(t_k)$ for the open-loop case, and uncorrelated with the reference signal $r(t_k)$ for the closed-loop case. For both settings, we assume for simplicity that the input signal $u(t)$ is known to be constant between samples, i.e., it has a zero-order hold (ZOH) behavior. Other known intersample behaviors, such as first-order hold or band-limited, can also be addressed within the proposed framework at the expense of a more involved notation and ad-hoc prefiltering techniques \cite{gonzalez2021srivc}. The closed-loop setting considers an input signal generated from a known discrete-time LTI controller $C_\textnormal{d}(q)$, with $q$ being the forward-shift operator. Formally, we can describe the input $u(t_k)$ in terms of the reference and output noise as follows:
\begin{equation}
\label{urv}
	u(t_k) = \tilde{r}(t_k)-\tilde{v}(t_k),
\end{equation}
where 
\begin{equation}
	\label{rtk}
	\tilde{r}(t_k):=S_{uo}^*(q) r(t_k), \hspace{0.2cm} 	\tilde{v}(t_k):=S_{uo}^*(q) v(t_k),
\end{equation}
with $S_{uo}^*(q)$ being the sensitivity function $S_{uo}^*(q)= C_{\textnormal{d}}(q)/[1+G_\textnormal{d}^*(q)C_\textnormal{d}(q)]^{-1}$, and $G_\textnormal{d}^*(q)$ being the ZOH-equivalent discrete-time system of $G^*(p)=\sum_{i=1}^K G_i^*(p)$.

We are concerned with developing data-driven methods to obtain estimates of the parameter vector
\begin{equation}
\label{betastar}
\bm{\beta}^*:= \begin{bmatrix}
\bm{\theta}_1^{*\top}, &\hspace{-0.05cm} \bm{\theta}_2^{*\top}, & \hspace{-0.05cm}\dots, & \hspace{-0.05cm}\bm{\theta}_K^{*\top}
\end{bmatrix}^\top. \notag
\end{equation}
To this end, we consider as given the input and output data $\{u(t_k),y(t_k)\}_{k=1}^N$ for the open-loop scenario, while $\{r(t_k)\}_{k=1}^N$ is also known in the closed-loop setting. 

\begin{rem}
    In Section \ref{sec:parsimonious} we also consider the identification of additive systems that are marginally stable. Since the denominator polynomials related to these systems cannot be represented in an anti-monic form as in \eqref{parametrization}, we provide the required system and model assumptions for this case separately.
\end{rem}

\begin{rem}
 Throughout this paper we assume that the input (for the open-loop case) and reference signals (for the closed-loop case) are quasi-stationary. Consequently, we will use the standard definition of expectation for quasi-stationary signals \cite[p. 34]{ljung1998system}
    \begin{equation}
        \overline{\mathbb{E}}\{g(t_k)\}:= \lim_{N\to\infty} \frac{1}{N}\sum_{k=1}^N \mathbb{E}\{g(t_k)\}. \notag
    \end{equation}
\end{rem}

\section{Stationary Points for Additive Continuous-time System Identification}
\label{sec:optimality}
In this section, we present the optimality conditions that will be exploited for designing the proposed additive system identification method in Section \ref{sec:parsimonious}. 

\subsection{Open-loop setting}
\label{subsec:openloop}
In the open-loop case, we seek an estimator that minimizes the output-error cost
\begin{equation}
	\label{betaol}
	\hat{\bm{\beta}}\hspace{-0.03cm}= \hspace{-0.03cm}\underset{\bm{\beta} \in \Omega}{\arg \min} \frac{1}{2N}\sum_{k=1}^N \hspace{-0.02cm}\left(y(t_k)\hspace{-0.04cm}-\hspace{-0.04cm}\sum_{i=1}^K G_i(p,\bm{\theta}_i)u(t_k)\hspace{-0.02cm}\right)^{\hspace{-0.03cm}2},
\end{equation}
where each $G_i(p,\bm{\theta}_i)$ is a transfer function parameterized as in \eqref{parametrization}. Note that the notation $G(p)u(t_k)$ means that the sampled input is interpolated with a zero-order hold, filtered through the continuous-time system $G(p)$, at later evaluated at $t=t_k$. The optimizer of the cost function in \eqref{betaol} must satisfy the first-order optimality condition
\begin{equation}
	\label{optimality}
	\frac{1}{N}\sum_{k=1}^N \hat{\bm{\varphi}}(t_k,\hat{\bm{\beta}}) \left(y(t_k)-\sum_{i=1}^K G_i(p,\hat{\bm{\theta}}_i)u(t_k)\right) = \mathbf{0},
\end{equation}
where $\hat{\bm{\beta}}$ is given by \eqref{betaol}, and the gradient of the residual can be written as
\begin{equation}
	\label{instrumentvector}
	\hat{\bm{\varphi}}(t_k,\hat{\bm{\beta}}) = \begin{bmatrix}
		\hat{\bm{\varphi}}_1^\top(t_k,\hat{\bm{\beta}}), &\hspace{-0.02cm} \dots, & \hspace{-0.02cm}\hat{\bm{\varphi}}_K^\top(t_k,\hat{\bm{\beta}})
	\end{bmatrix}^\top,
\end{equation}
with each vector $\hat{\bm{\varphi}}_i(t_k,\hat{\bm{\beta}})$ being given by
\begin{equation}
	\label{instrumentol}
\begin{split}
     &\hspace{-0.23cm}\hat{\bm{\varphi}}_i(t_k,\hat{\bm{\beta}}) = \bigg[\frac{-p\hat{B}_i(p)}{[\hat{A}_i(p)]^2} u(t_k), \dots, \frac{-p^{n_i} \hat{B}_i(p)}{[\hat{A}_i(p)]^2} u(t_k), \\
	&\hspace{1.7cm} \frac{1}{\hat{A}_i(p)}u(t_k),\dots, \frac{p^{m_i}}{\hat{A}_i(p)} u(t_k) \bigg]^{\top}. \hspace{-0.1cm}
\end{split}
\end{equation}

\subsection{Closed-loop setting}
In contrast to the open-loop analysis, an output-error loss function of the form in \eqref{betaol} applied to closed-loop system identification can result in an asymptotically biased estimator. This phenomenon is due to the correlation of the output noise with the input $u(t)$ through the closed-loop interconnection \cite{van1998closed,gonzalez2022consistency}. The bias in closed loop can be mitigated or reduced completely as $N$ tends to infinity if an instrumental variable approach is considered, as in \cite{gilson2011optimal} for the unfactored transfer function case. Thus, instead of \eqref{betaol}, for the closed-loop setting we are interested in computing the instrumental variable solution
\begin{equation}
	\label{betacl}
	\hat{\bm{\beta}}\in \underset{\bm{\beta} \in \Omega}{\textnormal{sol}} \hspace{-0.02cm} \left\{\hspace{-0.03cm}\frac{1}{N}\sum\limits_{k=1}^N \hspace{-0.07cm}\bm{\zeta}(\hspace{-0.01cm}t_k\hspace{-0.01cm})\hspace{-0.08cm}\left(\hspace{-0.05cm}y(\hspace{-0.01cm}t_k\hspace{-0.01cm})\hspace{-0.07cm}-\hspace{-0.05cm}\sum_{i=1}^K \hspace{-0.05cm} G_i\hspace{-0.02cm}(p,\hspace{-0.02cm}\bm{\theta}_i) u(t_k)\hspace{-0.03cm}\right)\hspace{-0.05cm} =\hspace{-0.03cm} \mathbf{0}\right\}\hspace{-0.02cm},
\end{equation}
where $\bm{\zeta}(t_k)\in\mathbb{R}^{K+\sum_{i=1}^K (n_i+m_i)}$ is an instrument vector that is assumed to be uncorrelated with the output noise $v(t_k)$. Such assumption is not restrictive in practice, since the instrument vector is typically formed by filtered versions of the reference signal, which are designed by the user.

A crucial aspect of the estimator in \eqref{betacl} is how the instrument vector is chosen for yielding consistent estimates of minimum covariance. Conditions on the instrument vector under which the estimator is generically consistent are provided in Lemma \ref{lemmaconsistency}.

\begin{lemma}
	\label{lemmaconsistency}
	Assume that the instrument vector $\bm{\zeta}(t_k)$ is uncorrelated with the output noise $v(t_k)$, and that $r(t_k)$ and $v(t_l)$ are quasi-stationary and mutually independent for all integers $k$ and $l$. Then, the estimator \eqref{betacl} is generically consistent if $\overline{\mathbb{E}}\left\{\bm{\zeta}(t_k) \tilde{\bm{\varphi}}^{r\top}(t_k)\right\}$ is generically nonsingular\footnote{In this context, a statement $s$, which depends on the elements $\bm{\rho}$ of some open set $\Omega\subseteq\mathbb{R}^n$, is \textit{generically true with respect to }$\Omega$ \cite{soderstrom1983instrumental} if the set $M=\{\bm{\rho}\in \Omega|s(\bm{\rho}) \textnormal{ is not true}\}$ has Lebesgue measure zero in $\Omega$.} with respect to the system and model denominator parameters, where $\tilde{\bm{\varphi}}^r(t_k)$ is formed by stacking the vectors $\{\tilde{\bm{\varphi}}_i^r(t_k)\}_{i=1}^K$, where
	\begin{equation}
		\label{alternativephi}
\begin{split}    
  \tilde{\bm{\varphi}}_i^r\hspace{-0.02cm}(t_k) \hspace{-0.05cm}= &\bigg[\frac{-pB_{i}^*(p)}{A_i\hspace{-0.02cm}(p)A_{i}^*\hspace{-0.02cm}(p)}\tilde{r}(t_k),\dots, \hspace{-0.02cm}\frac{-p^{n_i}B_i^*(p)}{A_i\hspace{-0.02cm}(p)A_i^*\hspace{-0.02cm}(p)}\tilde{r}(t_k), \\
		&\hspace{0.2cm}\frac{1}{A_{i}(p)}\tilde{r}(t_k), \dots, \frac{p^{m_i}}{A_{i}(p)}\tilde{r}(t_k)\bigg]^\top.
\end{split}
	\end{equation}
\end{lemma}
\textit{Proof.} As the sample size tends to infinity, the sum in $k$ in \eqref{betacl} converges to an expected value due to the quasi-stationarity assumptions on the reference and noise signals \cite{soderstrom1975ergodicity}, which leads to $\bar{\bm{\theta}}_i = \lim_{N\to \infty} \hat{\bm{\theta}}_i$ satisfying
\begin{align}
	\overline{\mathbb{E}}\left\{ \hspace{-0.02cm}\bm{\zeta}(\hspace{-0.01cm}t_k\hspace{-0.01cm})\hspace{-0.03cm}\left(y(\hspace{-0.01cm}t_k\hspace{-0.01cm})\hspace{-0.07cm}-\hspace{-0.05cm}\sum_{i=1}^K \hspace{-0.02cm} G_i\hspace{-0.01cm}(p,\hspace{-0.02cm}\bar{\bm{\theta}}_i) u(t_k)\hspace{-0.03cm}\right)\right\} &= \mathbf{0} \notag \\
	\iff \sum_{i=1}^K\overline{\mathbb{E}}\big\{ \hspace{-0.03cm}\bm{\zeta}(\hspace{-0.01cm}t_k\hspace{-0.01cm})\hspace{-0.04cm} \left[G_i^*\hspace{-0.02cm}(p)\hspace{-0.07cm}-\hspace{-0.07cm} G_i\hspace{-0.02cm}(p,\hspace{-0.02cm}\bar{\bm{\theta}}_i)\right]\hspace{-0.03cm} u(t_k)\hspace{-0.03cm}\big\} &= \mathbf{0} \notag \\
	\label{reducestocondition}
	\iff \sum_{i=1}^K\overline{\mathbb{E}}\big\{ \hspace{-0.03cm}\bm{\zeta}(\hspace{-0.01cm}t_k\hspace{-0.01cm})\hspace{-0.04cm} \left[G_i^*\hspace{-0.02cm}(p)\hspace{-0.07cm}-\hspace{-0.07cm} G_i\hspace{-0.02cm}(p,\hspace{-0.02cm}\bar{\bm{\theta}}_i)\right]\hspace{-0.03cm} \tilde{r}(t_k)\big\} &= \mathbf{0},
\end{align}
where we have used the fact that the instrument vector is uncorrelated with the output noise $v(t_k)$ in both steps (recall that $\tilde{r}(t_k)$ is given by \eqref{rtk}). Furthermore, we have
\begin{equation}
	\left[G_i^*\hspace{-0.02cm}(p)\hspace{-0.07cm}-\hspace{-0.07cm} G_i\hspace{-0.02cm}(p,\hspace{-0.02cm}\bar{\bm{\theta}}_i)\right]\hspace{-0.04cm} \tilde{r}(t_k) \hspace{-0.07cm}= \hspace{-0.07cm}\frac{1}{A_i^*\hspace{-0.02cm}(p)\bar{A}_i(p)}[1,p,\dots,p^{n_i\hspace{-0.01cm}+\hspace{-0.01cm}m_i}] \bm{\eta}_i, \notag
\end{equation}
where $\bm{\eta}_i\in \mathbb{R}^{n_i+m_i+1}$ is the vector that contains the coefficients of $A_i^*(p)\bar{B}_i(p)-\bar{A}_i(p)B_i^*(p)$ in descending order of degree. By following the derivation of Eq. (15) of \cite{pan2020consistency}, we find that
\begin{equation}
    \frac{1}{A_i^*\hspace{-0.02cm}(p)\bar{A}_i(p)}\mathbf{S}(-B_i^*,\hspace{-0.02cm}A_i^*)[1,p,\dots\hspace{-0.02cm},p^{n_i+m_i}]^{\hspace{-0.02cm}\top} \tilde{r}(t_k)\hspace{-0.05cm}=\hspace{-0.04cm}\tilde{\bm{\varphi}}_i^r(t_k), \notag
\end{equation}
where $\tilde{\bm{\varphi}}_i^r(t_k)$ is given in \eqref{alternativephi}, and $\mathbf{S}(-B_i^*,A_i^*)$ is a Sylvester matrix that is nonsingular for $i=1,\dots,K$ due to the coprimeness of the polynomials $A_i^*(p)$ and $B_i^*(p)$ \cite[Lemma A3.1]{soderstrom1983instrumental}. If the vectors $\tilde{\bm{\varphi}}^r(t_k)$ and $\bm{\eta}$ are formed by stacking the vectors $\tilde{\bm{\varphi}}_i^r(t_k)$ and $\bm{\eta}_i$ respectively, then \eqref{reducestocondition} reduces to the condition
\begin{equation}
	\overline{\mathbb{E}}\hspace{-0.03cm}\left\{\bm{\zeta}(t_k) \tilde{\bm{\varphi}}^{r\top}\hspace{-0.04cm}(t_k)\hspace{-0.03cm}\right\} \hspace{-0.07cm}\begin{bmatrix}
	\mathbf{S}(\hspace{-0.02cm}-\hspace{-0.02cm}B_1^*,\hspace{-0.02cm}A_1^*) & & \hspace{-0.15cm}\mathbf{0} \\ 
	 & \hspace{-0.15cm}\ddots &  \\ 
	\mathbf{0} & & \hspace{-0.15cm}\mathbf{S}(\hspace{-0.02cm}-\hspace{-0.02cm}B_{\hspace{-0.03cm}K}^*,\hspace{-0.02cm}A_{\hspace{-0.02cm}K}^*) \\ 
	\end{bmatrix}^{\hspace{-0.07cm}-\hspace{-0.02cm}\top}\hspace{-0.15cm}\bm{\eta} \hspace{-0.05cm}=\hspace{-0.05cm} \mathbf{0}, \notag 
\end{equation}
which implies that $\bm{\eta}$ is generically the zero vector (i.e., $G_i(p,\bar{\bm{\theta}}_i)=G_i^*(p)$ for all $i=1,\dots,K$) if the matrix $\overline{\mathbb{E}}\{\bm{\zeta}(t_k) \tilde{\bm{\varphi}}^{r\top}(t_k)\}$ is generically nonsingular. \hfill $\square$

\begin{rem}
The nonsingularity condition in Lemma \ref{lemmaconsistency} suggests that the instrument vector should be correlated with filtered versions of the derivatives of the noiseless model output and input. Such interpretation is ubiquitous in instrumental variable estimation, and has been thoroughly studied for the unfactored transfer function case \cite{Garnier2008book,boeren2018optimal,pan2020efficiency,mooren2023online}. Lemma \ref{lemmaconsistency} presents the first extension of these analyses to the additive system identification framework. 
\end{rem}

In accordance to Lemma \ref{lemmaconsistency}, we will focus solely on instrument vectors that meet the nonsingularity requirement, which defines a family of generically consistent estimators. Among these, we aim to characterize the one that obtains the least asymptotic covariance.  In cases where the transfer function is unfactored, it is established in both discrete-time \cite{stoica1983optimal,soderstrom1987instrumental} and continuous-time \cite{gilson2005instrumental} formulations that the instrument vector achieving the lower bound of the asymptotic covariance matrix for the parameter estimator is given by the noise-free component of the regressor vector. However, for the additive model description, the computation of $\hat{\bm{\beta}}$ cannot be carried out in a straightforward analytical manner, and the model output cannot be expressed using just one regressor vector. Despite these difficulties, Theorem \ref{thmnormality} confirms that the asymptotic distribution of $\hat{\bm{\beta}}$ in Eq. \eqref{betacl} follows a normal distribution, and the asymptotic covariance can be explicitly computed.

\begin{theorem}
	\label{thmnormality}
	Consider the estimator \eqref{betacl}. Assume that $r(t_k)$ and $v(t_l)$ are quasi-stationary and mutually independent for all integers $k$ and $l$, and that the instrument vector $\bm{\zeta}(t_k)$ is constructed such that $\hat{\bm{\beta}}$ is a consistent estimator of $\bm{\beta}^*$. Then, $\hat{\bm{\beta}}$ is asymptotically Gaussian distributed, i.e.,
	\begin{equation}
		\sqrt{N}(\hat{\bm{\beta}}-\bm{\beta}^*) \xrightarrow{\text{dist.}} \mathcal{N}(\mathbf{0},\mathbf{P}_{\textnormal{IV}}),
	\end{equation}
	where the asymptotic covariance matrix $\mathbf{P}_{\textnormal{IV}}$ is given by
	\begin{align}
		\mathbf{P}_{\textnormal{IV}}&=\sigma^2 \overline{\mathbb{E}}\left\{\bm{\zeta}(t_k)\bm{\psi}^\top(t_k)\right\}^{-1} \notag \\
		&\times \overline{\mathbb{E}}\left\{\bm{\zeta}(t_k)\bm{\zeta}^\top(t_k)\right\} \overline{\mathbb{E}}\left\{\bm{\psi}(t_k) \bm{\zeta}^\top(t_k)\right\}^{-1}, \notag 
	\end{align}
	where
	\begin{equation}
		\label{psi}
		\bm{\psi}(t_k) \hspace{-0.06cm}=\hspace{-0.08cm} \left[\hspace{-0.01cm}\frac{\partial G_1\hspace{-0.03cm}(p,\hspace{-0.02cm}\bm{\theta}_1\hspace{-0.02cm})}{\partial \bm{\theta}_1^\top},\hspace{-0.01cm}\dots\hspace{-0.01cm},\hspace{-0.03cm} \frac{\partial G_{\hspace{-0.02cm}K}\hspace{-0.02cm}(p,\hspace{-0.02cm}\bm{\theta}_K\hspace{-0.02cm})}{\partial \bm{\theta}_K^\top} \right]^{\hspace{-0.05cm}\top} \hspace{-0.06cm}\bigg|_{\bm{\beta}=\bm{\beta}^*} \hspace{-0.1cm}\tilde{r}(t_k)\hspace{0.02cm}.  \hspace{-0.09cm}
	\end{equation}
	
\end{theorem}
\textit{Proof.} A first-order Taylor expansion of the left hand side of \eqref{betacl} gives
\begin{align}
	\frac{1}{\sqrt{N}}& \sum_{k=1}^N \bm{\zeta}(t_k) \left(y(t_k)-\sum_{i=1}^K G_i^*(p) u(t_k)\right) \notag \\
	-&\frac{1}{\sqrt{N}} \sum_{k=1}^N \bm{\zeta}(t_k) \bm{\psi}^\top(t_k)(\hat{\bm{\beta}}-\bm{\beta}^*) \hspace{0.06cm}\# \hspace{0.06cm}\mathbf{0}, \notag 
\end{align}
where the $\#$ symbol means that the left-hand side and the right-hand side differ by a quantity converging to zero in probability, and where $\bm{\psi}(t_k)$ is defined in \eqref{psi}. Hence,
\begin{align}
	\hspace{-0.3cm}\sqrt{\hspace{-0.02cm}N}(\hat{\bm{\beta}}\hspace{-0.05cm}-\hspace{-0.05cm}\bm{\beta}^*\hspace{-0.03cm})&\# \hspace{-0.07cm}\left[\hspace{-0.02cm}\frac{1}{N}\hspace{-0.07cm}\sum_{k=1}^N \hspace{-0.04cm}\bm{\zeta}(t_k)\bm{\psi}^\top\hspace{-0.03cm}(t_k)\right]^{\hspace{-0.05cm}-\hspace{-0.03cm}1}\hspace{-0.12cm}\frac{1}{\sqrt{N}} \hspace{-0.03cm}\sum_{k=1}^N \hspace{-0.03cm}\bm{\zeta}(t_k) v(t_k) \notag \\
	\label{intermediate}
	&\# \hspace{0.05cm} \overline{\mathbb{E}}\hspace{-0.04cm}\left\{\hspace{-0.03cm}\bm{\zeta}(t_k)\bm{\psi}^{\hspace{-0.03cm}\top}\hspace{-0.05cm}(t_k)\hspace{-0.02cm}\right\}^{\hspace{-0.05cm}-1}\hspace{-0.12cm}\frac{1}{\sqrt{\hspace{-0.02cm}N}} \hspace{-0.04cm}\sum_{k=1}^N \hspace{-0.03cm}\bm{\zeta}(t_k) v(t_k). \hspace{-0.1cm}
\end{align}
It follows from Lemma A4.1 of \cite{soderstrom1983instrumental} that
\begin{equation}
\label{distconvergence}
	\frac{1}{\sqrt{N}} \sum_{k=1}^N \bm{\zeta}(t_k) v(t_k) \xrightarrow{\text{dist.}} \mathcal{N}(\mathbf{0},\mathbf{P}),
\end{equation}
where, following Eq. (22) of \cite{pan2020efficiency} and exploiting the fact that $\bm{\zeta}(t_k)$ and $v(t_k)$ are uncorrelated, we have
\begin{align}
	\mathbf{P}&= \lim_{N\to \infty}\frac{1}{N}\sum_{k=1}^N \sum_{s=1}^N \mathbb{E}\left\{[\bm{\zeta}(t_k)v(t_k)][\bm{\zeta}(t_s)v(t_s)]^\top \right\} \notag \\
	&= \sigma^2 \overline{\mathbb{E}}\left\{\bm{\zeta}(t_k)\bm{\zeta}^\top(t_k)\right\}. \notag
\end{align}
The desired result is then obtained from Lemma A4.2 of \cite{soderstrom1983instrumental} applied to \eqref{intermediate}. \hfill $\square$ 
 
\begin{corollary}
	\label{corocov}
	The asymptotic covariance matrix of $\hat{\bm{\beta}}$ in \eqref{betacl} is minimized in a positive definite sense when $\bm{\zeta}(t_k)=\bm{\psi}(t_k)$.	
\end{corollary}
\textit{Proof.} \label{proofcorocov}
Since any covariance matrix is positive semi-definite, we have
\begin{equation}
\label{covariancepositivedefinite}
	\overline{\mathbb{E}}\left\{\begin{bmatrix} \bm{\zeta}(t_k) \\
		\bm{\psi}(t_k)
	\end{bmatrix}\begin{bmatrix} \bm{\zeta}(t_k) \\
		\bm{\psi}(t_k)
	\end{bmatrix}^\top \right\} \succeq \mathbf{0}.
\end{equation}
The nonsingularity of $\overline{\mathbb{E}}\left\{\bm{\zeta}(t_k)\bm{\psi}^\top(t_k)\right\}$ implies that the matrix $\overline{\mathbb{E}}\left\{\bm{\zeta}(t_k)\bm{\zeta}^\top(t_k)\right\}$ is positive definite. Thus, \eqref{covariancepositivedefinite} is equivalent to the Schur complement
\begin{equation}
	\label{schurcomplement}
\begin{split}
 &\overline{\mathbb{E}}\left\{\bm{\psi}(t_k) \bm{\psi}^\top(t_k) \right\}-\overline{\mathbb{E}}\left\{\bm{\psi}(t_k) \bm{\zeta}^\top(t_k)\right\} \\ 
	&\hspace{0.1cm}\times \overline{\mathbb{E}}\left\{\bm{\zeta}(t_k)\bm{\zeta}^\top(t_k)\right\}^{-1}\overline{\mathbb{E}}\left\{\bm{\zeta}(t_k)\bm{\psi}^\top(t_k)\right\}\succeq \mathbf{0}.    
\end{split}
\end{equation}
Provided that the matrices above are nonsingular, \eqref{schurcomplement} is equivalent to
\begin{align}
	&\overline{\mathbb{E}}\left\{\bm{\psi}(t_k) \bm{\psi}^\top(t_k) \right\}^{-1}\preceq \overline{\mathbb{E}}\left\{\bm{\zeta}(t_k)\bm{\psi}^\top(t_k)\right\}^{-1} \notag \\
	&\hspace{0.7cm}\times \overline{\mathbb{E}}\left\{\bm{\zeta}(t_k)\bm{\zeta}^\top(t_k)\right\}\overline{\mathbb{E}}\left\{\bm{\psi}(t_k) \bm{\zeta}^\top(t_k)\right\}^{-1}. \notag
\end{align}
Equality holds when $\bm{\zeta}(t_k)=\mathbf{M}\bm{\psi}(t_k)$, with $\mathbf{M}$ being a nonsingular constant matrix. Since $\mathbf{M}$ can be canceled out from \eqref{betacl}, we set $\mathbf{M}$ equal to the identity matrix without loss of generality, which leads to the desired result. \hfill $\square$ 				
 
Corollary \ref{corocov} indicates that the optimal instrument vector is given by a filtered version of the reference signal that depends on the true system parameters, as seen in \eqref{psi}. Since the true parameters are not known, the solution for the unfactored transfer function case adopted in, e.g., the closed-loop SRIVC estimator (CLSRIVC, \cite{gilson2008instrumental}), is to let $\bm{\zeta}(t_k)$ depend on $\bm{\beta}$ as well. In the additive identification case, we let the instrument vector take the form \eqref{instrumentvector}, but where, for $i=1,\dots,K$,
\begin{equation}
    \begin{split}
	\label{instrumentcl}
	\bm{\zeta}_i&(t_k,\bm{\beta}) \hspace{-0.06cm}= \hspace{-0.06cm}\bigg[\hspace{-0.03cm}\frac{-pB_{i}(p)}{[A_{i}(p)]^2},\dots, \hspace{-0.03cm}\frac{-p^{n_i}B_{i}(p)}{[A_{i}(p)]^2}, \\
	&\hspace{1.5cm}\frac{1}{A_{i}(p)}, \dots, \frac{p^{m_i}}{A_{i}(p)}\bigg]^\top S_{uo}(q) r(t_k),        
    \end{split}
\end{equation}
where $S_{uo}(q) = C_{\textnormal{d}}(q)/[1+G_\textnormal{d}(q)C_\textnormal{d}(q)]^{-1}$. 

In summary, the open-loop and closed-loop identification problems can be solved using the optimality condition \eqref{optimality} and the (refined) instrumental variable expression \eqref{betacl}, respectively. In the closed-loop scenario, the instrument $\bm{\zeta}(t_k)$ depends on the model, as indicated in~\eqref{instrumentcl}.

\begin{rem}
    To unify the notations pertaining to the open and closed loop cases, from now on we denote the instrument vector composed by stacking the vectors \eqref{instrumentcl} for $i=1,\dots,K$ as $\hat{\bm{\varphi}}(t_k,\bm{\beta})$. The difference between this vector and the one described by \eqref{instrumentol} will be clear from the context.
\end{rem}

\section{Additive System Identification: An Instrumental Variable Solution}
\label{sec:parsimonious}
This section presents the proposed instrumental variable-based method for identifying additive continuous-time systems in open or closed loop settings, and its consistency analysis.

\subsection{Derivation of the instrumental variable solution}
When computing $\hat{\bm{\beta}}$ that satisfies \eqref{optimality} or \eqref{betacl}, the difference between the measured and predicted outputs cannot be written as a unique pseudolinear regression when $K>1$ due to the additive structure of the model. Then, to estimate additive models, we write the residual in $K$ different ways:
\begin{equation}
	\label{differentways0}
	y(t_k)-\sum_{i=1}^K G_i(p,\hat{\bm{\theta}}_i) u(t_k) = y_{f,i}(t_k,\hat{\bm{\theta}}_{i})-\bm{\varphi}_i^{\top}(t_k,\hat{\bm{\theta}}_{i})\hat{\bm{\theta}}_{i} ,
\end{equation}
where $i = 1,\dots, K$. Here, $y_{f,i}(t_k,\hat{\bm{\beta}})$ represents a filtered \textit{residual} output of the form
\begin{equation}
\label{residualoutputyfi}
    y_{f,i}(t_k,\hat{\bm{\beta}}) = \frac{1}{\hat{A}_i(p)} \tilde{y}_i(t_k),
\end{equation}
where the residual output pertaining to the $i$th additive submodel is given by
\begin{equation}
    \tilde{y}_i(t_k) = y(t_k)-\sum_{\substack{l = 1,\dots,K,\\ l \neq i}} G_l(p,\hat{\bm{\theta}}_l)u(t_k),
\end{equation}
and the regressor vector associated with the $i$th additive submodel, denoted by $\bm{\varphi}_{i}(t_k,\hat{\bm{\theta}}_i)$, is expressed as
\begin{equation}
\begin{split}
    \label{filteredregressorresidual}
	    &\hspace{-0.23cm}\bm{\varphi}_{i}(t_k,\hat{\bm{\theta}}_i) = \bigg[\frac{-p}{\hat{A}_i(p)} \tilde{y}_i(t_k), \dots, \frac{-p^{n_i}}{\hat{A}_i(p)} \tilde{y}_i(t_k), \\
	&\hspace{1.7cm} \frac{1}{\hat{A}_i(p)}u(t_k),\dots, \frac{p^{m_i}}{\hat{A}_i(p)} u(t_k) \bigg]^{\top}.
\end{split}
\end{equation}
Thus, if we define the following signals and matrix
\begin{align}
	\label{fullregressor}
	\bm{\varphi}(t_k,\hat{\bm{\beta}}) &= [\bm{\varphi}_1^{\top}(t_k,\hat{\bm{\theta}}_{1}), \dots, \bm{\varphi}_K^{\top}(t_k,\hat{\bm{\theta}}_{K})]^\top, \\
	\label{fullfilteredoutput}
	\bm{\Upsilon}(t_k,\hat{\bm{\beta}}) &= [y_{f,1}(t_k,\hat{\bm{\beta}}), \dots, y_{f,K}(t_k,\hat{\bm{\beta}})]^\top,  \\
	\label{mathcalB}
	\hat{\mathcal{B}} &= \begin{bmatrix}
		\hat{\bm{\theta}}_1 & &\mathbf{0} \\
		& \ddots &  \\
		\mathbf{0} & & \hat{\bm{\theta}}_K 
	\end{bmatrix},
\end{align}
the conditions in \eqref{optimality} and \eqref{betacl} can be written in $K$ different ways using \eqref{differentways0}, which leads to the following equivalent equation that the optimal estimate must satisfy:
\begin{equation}
\label{leadingto}
	\frac{1}{N}\sum_{k=1}^N \hat{\bm{\varphi}}(t_k,\hat{\bm{\beta}}) \left(\bm{\Upsilon}^\top(t_k,\hat{\bm{\beta}})-\bm{\varphi}^\top(t_k,\hat{\bm{\beta}})\hat{\mathcal{B}} \right) = \mathbf{0}.
\end{equation}
Thus, by fixing $\hat{\bm{\beta}}=\hat{\bm{\beta}}^j$ in $\hat{\bm{\varphi}}(t_k,\hat{\bm{\beta}})$, $\bm{\Upsilon}^\top(t_k,\hat{\bm{\beta}})$ and $\bm{\varphi}^\top(t_k,\hat{\bm{\beta}})$, we obtain the $j$th iteration of the proposed estimator for both open and closed loop settings by solving for $\hat{\mathcal{B}}$ as follows:
\begin{equation}
\begin{split}
	\label{nonsingularity}
	\mathcal{B}^{j+1} &= \left[\frac{1}{N}\sum_{k=1}^N \hat{\bm{\varphi}}(t_k,\bm{\beta}^j)\bm{\varphi}^\top(t_k,\bm{\beta}^j) \right]^{-1}  \\
	&\hspace{0.4cm}\times \left[\frac{1}{N}\sum_{k=1}^N \hat{\bm{\varphi}}(t_k,\bm{\beta}^j)\bm{\Upsilon}^\top(t_k,\bm{\beta}^j) \right], 
\end{split}
\end{equation}
where the next iteration $\bm{\beta}^{j+1}$ is extracted from the block diagonal coefficients of $\mathcal{B}^{j+1}$ as in \eqref{mathcalB}, and the instrument vector is described by \eqref{instrumentol} or \eqref{instrumentcl} for the open or closed loop setups, respectively. The iterations in \eqref{nonsingularity} are computed until convergence for \eqref{optimality} or \eqref{betacl} to be satisfied.
\begin{rem}
    The iterations in \eqref{nonsingularity} can be viewed as an extension of the open-loop and closed-loop refined instrumental variable estimators to the additive model case. When we consider $K=1$, i.e., the unfactored transfer function case, the iterations in \eqref{nonsingularity} reduce to
\begin{equation}
\begin{split}
\label{srivc}
\bm{\theta}_1^{j+1} &= \left[\frac{1}{N}\sum_{k=1}^N \hat{\bm{\varphi}}_1(t_k,\bm{\theta}_1^j)\bm{\varphi}_1^\top(t_k,\bm{\theta}_1^j) \right]^{-1}  \\
\times &\left[\frac{1}{N}\sum_{k=1}^N \hat{\bm{\varphi}}_1(t_k,\bm{\theta}_1^j)y_{f,1}(t_k,\bm{\theta}_1^j) \right], 
\end{split}
\end{equation}
where the instrument vector $\hat{\bm{\varphi}}_1(t_k,\bm{\theta}_1^j)$ is given by \eqref{instrumentol} for the open-loop case, and \eqref{instrumentcl} for the closed-loop setting. These iterations correspond to the standard SRIVC and CLSRIVC methods, respectively \cite{young1980refined,gilson2008instrumental}. The SRIVC and CLSRIVC estimators have recently been proven to be generically consistent under mild conditions \cite{pan2020consistency,gonzalez2022consistency}, and the asymptotic efficiency of the SRIVC estimator has been proven in \cite{pan2020efficiency}. 
\end{rem}
\begin{rem}
\label{rem:marginally}
    The proposed approach can be extended to the estimation of marginally stable systems as follows. Consider the system \eqref{ctsystem1}, but where the first submodel has $\ell$ poles at the origin. In other words, let $G^*_1(p) = B_1^*(p)/[p^\ell A_1^*(p)]$, where $A_{1}^*(p)$ and $B_1^*(p)$ are coprime polynomials, and have the same form as in \eqref{parametrization}. For either open or closed-loop variants of the proposed approach, we require the computation of the gradient of each submodel with respect to their respective parameter vector. For $G_1(p)$, this computation leads to the following instrument vector:
\begin{equation}
\label{instrument_int}
    \begin{split}
	\hat{\bm{\varphi}}_1&(t_k,\bm{\beta}) \hspace{-0.06cm}= \hspace{-0.06cm}\bigg[\hspace{-0.03cm}\frac{-pB_{1}(p)}{p^{\ell}[A_{1}(p)]^2},\dots, \hspace{-0.03cm}\frac{-p^{n_1}B_{1}(p)}{p^{\ell}[A_{1}(p)]^2}, \\
	&\hspace{1.5cm}\frac{1}{p^{\ell}A_{1}(p)}, \dots, \frac{p^{m_1}}{p^{\ell}A_{1}(p)}\bigg]^\top z(t_k),        
    \end{split}
\end{equation}
where $z(t_k)=u(t_k)$ for the open-loop algorithm, and $z(t_k)=S_{uo}(q) r(t_k)$ for the closed-loop variant. On the other hand, the model residual retains the same form as in \eqref{differentways0}, with the filtered residual output given by \eqref{residualoutputyfi}, but with the regressor vector now expressed as
\begin{equation}
\label{regressor_int}
\begin{split}
    &\hspace{-0.23cm}\bm{\varphi}_{1}(t_k,\hat{\bm{\theta}}_1) = \bigg[\frac{-p}{\hat{A}_1(p)} \tilde{y}_1(t_k), \dots, \frac{-p^{n_1}}{\hat{A}_1(p)} \tilde{y}_1(t_k), \\
	&\hspace{1.7cm} \frac{1}{p^\ell \hat{A}_1(p)}u(t_k),\dots, \frac{p^{m_1}}{p^\ell\hat{A}_1(p)} u(t_k) \bigg]^{\top}.
 \end{split}
\end{equation}
By computing the iterations in \eqref{nonsingularity} with the first block of the instrument and regressor vectors given by \eqref{instrument_int} and \eqref{regressor_int} respectively, we obtain a direct extension of the proposed method for identifying marginally stable systems in additive form. This solution extends the one proposed in Solution~6.2 of \cite{gonzalez2022unstable}, which is only applicable to unfactored transfer functions in closed loop. We note that the user's choice of including an integral action term in the model can be motivated by physical intuition, or by model order selection methods \cite{ljung1998system}.
\end{rem}

Note that a block diagonal structure for $\mathcal{B}^{j+1}$ is not achieved in general at each iteration \eqref{nonsingularity}, but it is guaranteed if the iterations converge to a stationary point that satisfies the pseudolinear regression equations in \eqref{leadingto}. This result is presented in Lemma \ref{lemmablockdiagonal}.

\begin{lemma}
    \label{lemmablockdiagonal} Consider the iterative procedure in \eqref{nonsingularity} for finite $N$ with $\bm{\varphi}(t_k,\bm{\beta}^j)$ and $\bm{\Upsilon}(t_k,\bm{\beta}^j)$ given by \eqref{fullregressor} and \eqref{fullfilteredoutput} respectively, while $\hat{\bm{\varphi}}(t_k,\bm{\beta}^j)$ is formed by \eqref{instrumentol} or \eqref{instrumentcl} in the open and closed-loop setups, respectively. Denote $\bar{\bm{\beta}}$ as the converging point of the iterative procedure, with $\bar{\mathcal{B}}=\lim_{j\to \infty}\mathcal{B}^{j}$, and assume that the matrix
    \begin{equation}
    \label{modifiednormalmatrix}
        \left[\frac{1}{N}\sum_{k=1}^N \hat{\bm{\varphi}}(t_k,\bar{\bm{\beta}})\bm{\varphi}^\top(t_k,\bar{\bm{\beta}}) \right]
    \end{equation}
    is nonsingular. Then, the converging point of the iterative procedure satisfies \eqref{optimality} or \eqref{betacl} if and only if the matrix $\bar{\mathcal{B}}$ is block-diagonal.    
\end{lemma}
\textit{Proof.} \label{prooflemmablockdiagonal}
The straight implication part of the result is direct from the derivation leading to \eqref{leadingto} and \eqref{nonsingularity}. For the converse, we note that the converging point of the iterative procedure in \eqref{nonsingularity} must satisfy
\begin{align}
	\bar{\mathcal{B}} \hspace{-0.03cm} =\hspace{-0.08cm} \left[\sum_{k=1}^N \hspace{-0.02cm}\hat{\bm{\varphi}}(t_k,\bar{\bm{\beta}})\bm{\varphi}^{\hspace{-0.04cm}\top}\hspace{-0.03cm}(t_k,\bar{\bm{\beta}}) \right]^{\hspace{-0.03cm}-1}\hspace{-0.04cm}\left[\sum_{k=1}^N \hat{\bm{\varphi}}(t_k,\bar{\bm{\beta}})\bm{\Upsilon}^\top(t_k,\bar{\bm{\beta}}) \right]. \notag 
\end{align}
Equivalently,
\begin{equation}
    \sum_{k=1}^N \hat{\bm{\varphi}}(t_k,\bar{\bm{\beta}})\left[\bm{\Upsilon}^\top(t_k,\bar{\bm{\beta}}) -\bm{\varphi}^\top(t_k,\bar{\bm{\beta}})\bar{\mathcal{B}}  \right] = \mathbf{0}. \notag
\end{equation}
Define $\tilde{\mathcal{B}}=\textnormal{diag}(\bar{\bm{\theta}}_{1},\dots,\bar{\bm{\theta}}_{K})$. This matrix satisfies
\begin{align}
    \bm{\Upsilon}^\top&(t_k,\bar{\bm{\beta}}) -\bm{\varphi}^\top(t_k,\bar{\bm{\beta}})\tilde{\mathcal{B}} \notag \\
    &=\left(y(t_k)-\sum_{i=1}^K G_i(p,\bar{\bm{\theta}}_{i}) u(t_k)\right)[1,\dots,1], \notag
\end{align}
which shows that the matrix formed by the off-diagonal elements of $\bar{\mathcal{B}}$, denoted as $\mathbf{E}=\bar{\mathcal{B}}-\tilde{\mathcal{B}}$, must satisfy
\begin{align}
    \sum_{k=1}^N &\hat{\bm{\varphi}}(t_k,\bar{\bm{\beta}})\bm{\varphi}^\top(t_k,\bar{\bm{\beta}})\mathbf{E} \notag \\
    =& \sum_{k=1}^N \hat{\bm{\varphi}}(t_k,\bar{\bm{\beta}})\hspace{-0.06cm}\left(y(t_k)\hspace{-0.06cm}-\hspace{-0.06cm}\sum_{i=1}^K G_i(p,\bar{\bm{\theta}}_{i}) u(t_k)\hspace{-0.03cm}\right)[1,\dots,1] \notag \\
    =& \hspace{0.05cm}\mathbf{0}, \notag 
\end{align}
where the last equality is due to the assumption that the limiting point $\bar{\bm{\beta}}$ satisfies the optimality condition \eqref{optimality} or \eqref{betacl} in the open or closed-loop setting, respectively. Since the modified normal matrix \eqref{modifiednormalmatrix} is assumed to be nonsingular, we reach $\mathbf{E} = \mathbf{0}$, which concludes the proof. \hfill $\square$ 

As seen in Lemma \ref{lemmablockdiagonal}, the nonsingularity of the matrix \eqref{nonsingularity} is essential for the method to be well posed. This aspect, as well as the convergence of the iterations, will be studied extensively in the next subsection.

\subsection{Consistency Analysis}
\label{sec:analysis}
Before presenting the main theorem on the generic consistency of the proposed method for open and closed loop, we shall introduce some signals of interest. By substituting \eqref{output} into \eqref{filteredregressorresidual}, we can express the regressor vector $\bm{\varphi}(t_k,\bm{\beta})$ as the block vector formed by
\begin{equation}
	\bm{\varphi}_i(t_k,\bm{\beta}) = \tilde{\bm{\varphi}}_i^u(t_k,\bm{\beta})+ \bm{\Delta}_i^u(t_k,\bm{\beta})-\textbf{v}_i^u(t_k,\bm{\beta}), \notag 
\end{equation}
where $\tilde{\bm{\varphi}}_i^u(t_k,\bm{\beta})$ has the same form as $\tilde{\bm{\varphi}}_i^r(t_k,\bm{\beta})$ in \eqref{alternativephi} but with $\tilde{r}(t_k)$ replaced by $u(t_k)$. That is,
\begin{align}
	&\hspace{-0.23cm}\tilde{\bm{\varphi}}_i^u(t_k,\bm{\beta}) = \bigg[\frac{-pB^*_i(p)}{A_i(p)A_i^*(p)} u(t_k), \dots, \frac{-p^{n_i} B^*_i(p)}{A_i(p)A_i^*(p)} u(t_k), \notag \\
    \label{errorfreeregressor}
	&\hspace{1.7cm} \frac{1}{A_i(p)}u(t_k),\dots, \frac{p^{m_i}}{A_i(p)} u(t_k) \bigg]^{\top}. \hspace{-0.1cm}
\end{align}
The vector $\bm{\Delta}^u(t_k,\bm{\beta})$ contains the interpolation errors that arise from constructing the filtered disturbance-free derivatives of the output, as well as the residual model bias. The entries of $\bm{\Delta}_i^u(t_k,\bm{\beta})$, denoted $\bm{\Delta}_{i,j}^u(t_k,\bm{\beta})$, are zero for $j>n_i$ and 
\begin{align}
	\bm{\Delta}_{i,j}^u\hspace{-0.01cm}(t_k,\hspace{-0.02cm}\bm{\beta})\hspace{-0.07cm} &=\hspace{-0.07cm} \frac{p^j B_i^*(p)}{A_i(p)A_i^*\hspace{-0.02cm}(p)}u(\hspace{-0.01cm}t_k\hspace{-0.01cm}) \hspace{-0.06cm}-\hspace{-0.06cm}\frac{p^j}{A_{i}\hspace{-0.01cm}(p)}\hspace{-0.07cm}\left\{\hspace{-0.05cm}\frac{B_i^*\hspace{-0.02cm}(p)}{A_i^*\hspace{-0.02cm}(p)}u(t)\hspace{-0.06cm}\right\}_{\hspace{-0.05cm}t=t_k} \notag \\
	\label{deltaij}
	&\hspace{-1cm}- \sum_{\substack{l = 1,\dots,K,\\ l \neq i}} \frac{p^j}{A_i(p)} \big\{(G_l^*(p)-G_l(p))u(t)\big\}_{t=t_k} 
\end{align}
for $j=1,\dots,n_i$. The direct contribution of the noise to the regressor is given by
\begin{equation}
\label{contributionnoise}
	\mathbf{v}_{i}^u(t_k,\bm{\beta})=\bigg[\frac{p}{A_{i}(p)}v(t_k), \dots, \frac{p^{n_i}}{A_{i}(p)}v(t_k), 0,\dots,0 \bigg]^\top. \notag
\end{equation}
In the closed-loop case, the input also contains filtered output noise that must be rewritten for the analysis. By exploiting \eqref{urv}, the regressor vector in the closed-loop setting can also be expressed as
\begin{equation}
	\bm{\varphi}_i(t_k,\bm{\beta}) = \tilde{\bm{\varphi}}_i^r(t_k,\bm{\beta})+ \bm{\Delta}_i^r(t_k,\bm{\beta})-\textbf{v}_i^r(t_k,\bm{\beta}), \notag 
\end{equation}
where $\tilde{\bm{\varphi}}_i^r(t_k,\bm{\beta})$ is given by \eqref{alternativephi}, and $\bm{\Delta}_i^r(t_k,\bm{\beta})$ has the same form as \eqref{deltaij} but with $\tilde{r}(t_k)$ instead of $u(t_k)$, and $\textbf{v}_i^r(t_k,\bm{\beta})$ solely contains filtered versions of the output noise $v(t_k)$. 

For the main result of this section, we require the following assumptions, some of which are divided into open or closed-loop assumptions depending on the setting we consider.
\begin{assumption}[Persistency of excitation]
	\label{assumption1}
	Open loop: The input $u(t_k)$ is persistently exciting of order no less than $2n$. Closed loop: The input $r(t_k)$ is persistently exciting of order no less than $2n+n_c$, where $n_c$ is the order of the discrete-time LTI controller $C_\textnormal{d}(q)$.
\end{assumption}
\begin{assumption}[Stationarity and independence]
	\label{assumption2}
	Open loop: The input $u(t_k)$ and disturbance $v(t_s)$ are quasi-stationary and mutually independent for all $k$ and $s$. Closed loop: The reference $r(t_k)$ and disturbance $v(t_s)$ are quasi-stationary and mutually independent for all $k$ and $s$.
\end{assumption}
\begin{assumption}[Stability and coprimeness]
	\label{assumption3}
	All the zeros of the $j$th iteration of the model denominator estimate $A_{i,j}(p)$ have strictly negative real parts, with $A_{i,j}(p)$ and $B_{i,j}(p)$ being coprime. Closed loop: Additionally, the discrete-time equivalent model estimates at each iteration are asymptotically stabilized by the controller $C_\textnormal{d}(q)$.
\end{assumption}
\begin{assumption}[Sampling rate]
	\label{assumption4}
	The sampling frequency is more than twice that of the largest imaginary part of the zeros of $\prod_{j=1}^K A_j(p)A_j^*(p)$.
\end{assumption}
Assumptions \ref{assumption1} to \ref{assumption4} are the same ones that have been considered in recent consistency analyses of open-loop and closed-loop refined instrumental variable estimators \cite{pan2020consistency,gonzalez2022consistency}. Assumptions \ref{assumption1} and \ref{assumption2} can be readily satisfied if the input is white noise or a periodic signal of sufficient number of frequency lines. The stability condition in Assumption~\ref{assumption3} is not restrictive, since the poles of the model at each iteration are typically reflected to the left half-plane if they are unstable. Cancelling the unstable poles using all-pass filters also circumvents this problem \cite{gonzalez2022unstable}. Finally, Assumption \ref{assumption4} avoids aliasing issues associated with the discrete-time implementation of the prefiltering steps, and is satisfied in practice if the sampling period is chosen small compared to the rise time of the system.

Theorem \ref{thmconsistency} shows that the proposed method in generically consistent for both open and closed-loop settings.

\begin{theorem}
	\label{thmconsistency}
	Consider the proposed estimator \eqref{nonsingularity} for the open and closed-loop settings in Figure \ref{fig_setups}, and suppose Assumptions \ref{assumption1} to \ref{assumption4} hold. Then, the following statements are true:
	\begin{enumerate}
		\item The modified normal matrix $\overline{\mathbb{E}}\left\{\hat{\bm{\varphi}}(t_k,\bm{\beta})\bm{\varphi}^{\top}(t_k,\bm{\beta}) \right\}$ is generically nonsingular with respect to the system and model denominator polynomials, provided that the following condition holds for each respective setting:
  \begin{enumerate}
      \item Open loop:
		\begin{equation}
			\label{normconditionol}
   \begin{split}
       \hspace{-0.85cm}\big\|&\overline{\mathbb{E}}\left\{\hat{\bm{\varphi}}(t_k,\bm{\beta})\bm{\Delta}^{u\top}(t_k,\bm{\beta}) \right\}\big\|_2  \\		
&<\sigma_{\textnormal{min}}\left(\overline{\mathbb{E}}\left\{\hat{\bm{\varphi}}(t_k,\bm{\beta})\tilde{\bm{\varphi}}^{u\top}(t_k,\bm{\beta}) \right\}\right),
\end{split}
		\end{equation}
	where $\sigma_{\textnormal{min}}(\cdot)$ represents the minimum singular value of a matrix, and $\hat{\bm{\varphi}}(t_k,\bm{\beta})$, $\tilde{\bm{\varphi}}^u(t_k,\bm{\beta})$ and $\bm{\Delta}^u(t_k,\bm{\beta})$ are formed by stacking the vectors described by \eqref{instrumentol}, \eqref{errorfreeregressor} and \eqref{deltaij}, respectively.
      \item Closed loop:
		\begin{equation}
  \label{normconditioncl}
  \begin{split}
			\hspace{-0.85cm}\big\|&\overline{\mathbb{E}}\left\{ \hat{\bm{\varphi}}(t_k,\bm{\beta})\bm{\Delta}^{r\top}(t_k,\bm{\beta})\right\} \big\|_2 \\
&<\sigma_{\textnormal{min}}\left(\overline{\mathbb{E}}\left\{\hat{\bm{\varphi}}(t_k,\bm{\beta})\tilde{\bm{\varphi}}^{r\top}(t_k,\bm{\beta}) \right\} \right),
		\end{split}
  \end{equation}
	where $\hat{\bm{\varphi}}(t_k,\bm{\beta})$, $\tilde{\bm{\varphi}}^r(t_k,\bm{\beta})$ are formed by stacking the vectors described by \eqref{instrumentcl} and \eqref{alternativephi}, respectively, and $\bm{\Delta}^r(t_k,\bm{\beta})$ has the same form as \eqref{deltaij} but with $\tilde{r}(t_k)$ instead of $u(t_k)$.
  \end{enumerate}	
		\item Assume that \eqref{normconditionol} for open loop or \eqref{normconditioncl} for closed loop is satisfied, and the iterations of the estimator converge for all $N$ sufficiently large to say, $\bar{\mathcal{B}}$, with $\bar{\mathcal{B}}$ being a block-diagonal matrix formed by $\bar{\bm{\beta}}_i$. Then, the true parameter vector $\bm{\beta}^*$ is the unique converging point of $\bar{\bm{\beta}}$ as the sample size tends to infinity.
 	\end{enumerate}
\end{theorem}
\begin{rem}
    The expressions in (36) and (37) give sufficient but not necessary conditions for the generic nonsingularity of $\overline{\mathbb{E}}\left\{\hat{\bm{\varphi}}(t_k,\bm{\beta})\bm{\varphi}^{\top}(t_k,\bm{\beta}) \right\}$. They are satisfied in practice when the respective interpolation error $\bm{\Delta}^u(t_k,\bm{\beta})$ or $\bm{\Delta}^r(t_k,\bm{\beta})$ arising from constructing the filtered output derivatives is not significant, which typically occurs when the sampling period is chosen appropriately.
\end{rem}

\textit{Proof of Statement 1, Part (a)}: Since the input is uncorrelated with the output noise, the expectation of interest is given by
\begin{equation}
\overline{\mathbb{E}}\hspace{-0.05cm} \left\{ \hspace{-0.035cm} \hat{\bm{\varphi}}(t_k,\hspace{-0.025cm} \bm{\beta}\hspace{-0.01cm})\bm{\varphi}^{\hspace{-0.06cm}\top}\hspace{-0.07cm} (\hspace{-0.01cm}t_k,\hspace{-0.02cm} \bm{\beta}\hspace{-0.01cm}) \hspace{-0.02cm}\right\} \hspace{-0.11cm} =\hspace{-0.07cm}  \overline{\mathbb{E}}\hspace{-0.05cm} \left\{ \hspace{-0.035cm} \hat{\bm{\varphi}}(\hspace{-0.01cm}t_k,\hspace{-0.02cm} \bm{\beta}\hspace{-0.01cm})[\tilde{\bm{\varphi}}^u\hspace{-0.03cm}(\hspace{-0.01cm}t_k,\hspace{-0.02cm} \bm{\beta}\hspace{-0.01cm})\hspace{-0.075cm}+\hspace{-0.08cm} \bm{\Delta}^{\hspace{-0.05cm}u}\hspace{-0.03cm}(\hspace{-0.01cm}t_k,\hspace{-0.02cm} \bm{\beta}\hspace{-0.01cm})]^{\hspace{-0.04cm}\top} \hspace{-0.03cm} \right\}\hspace{-0.07cm} . \notag
\end{equation}
Provided the condition \eqref{normconditionol} holds, due to Theorem 5.1 of \cite{dahleh2002lectures}, the perturbation matrix $\overline{\mathbb{E}}\left\{ \hat{\bm{\varphi}}(t_k,\bm{\beta})\bm{\Delta}^{u\top}(t_k,\bm{\beta}) \right\}$ is small enough (in 2-norm) to not affect the nonsingularity of the matrix $\overline{\mathbb{E}}\left\{ \hat{\bm{\varphi}}(t_k,\bm{\beta})\bm{\varphi}^\top(t_k,\bm{\beta}) \right\}$. Thus, we study the generic nonsingularity of the matrix $\overline{\mathbb{E}}\left\{ \hat{\bm{\varphi}}(t_k,\bm{\beta})\tilde{\bm{\varphi}}^{u\top}(t_k,\bm{\beta}) \right\}$ instead.

By following standard operations (see, e.g., \cite{gonzalez2022consistency}), the noise-free, interpolation-error-free regressor vector and the instrument vector can be rewritten using the identities 
\begin{align}
\tilde{\bm{\varphi}}_i^u(t_k,\bm{\beta})&= \mathbf{S}(-B_i^*,A_i^*) \frac{1}{A_i(p)A_i^*(p)} \mathbf{u}_{i}(t_k), \\
\hat{\bm{\varphi}}_i(t_k,\bm{\beta})&= \mathbf{S}(-B_i,A_i) \frac{1}{A_i^2(p)} \mathbf{u}_{i}(t_k),
\end{align}
where $\mathbf{S}(-B_i^*,A_i^*)$ and $\mathbf{S}(-B_i,A_i)$ are Sylvester matrices that are nonsingular due to the coprimeness of the polynomials of the true system $G_i^*(p)$ and the coprimeness assumption on the model $G_i(p)$ \cite[Lemma A3.1]{soderstrom1983instrumental}. The vector $\mathbf{u}_i(t_k)$ is formed by input derivatives, i.e.,
\begin{equation}
\label{derivativesu}
	\mathbf{u}_i(t_k) = \begin{bmatrix}
		p^{n_i+m_i}, & p^{n_i+m_i-1}, & \dots, &  1 
	\end{bmatrix}^\top u(t_k). 
\end{equation}
Hence, 
\begin{equation}
\label{inorder}
\overline{\mathbb{E}}\left\{ \hat{\bm{\varphi}}(t_k,\bm{\beta})\tilde{\bm{\varphi}}^{u\top}(t_k,\bm{\beta})\right\} = \boldsymbol{\mathcal{S}} \bm{\Phi}_u \boldsymbol{\mathcal{S}}_*^\top,
\end{equation} 
where
\begin{equation}
	\label{phimatrix}
	\bm{\Phi}_u \hspace{-0.07cm}=\hspace{-0.06cm} \overline{\mathbb{E}}\hspace{-0.05cm}\left\{\hspace{-0.05cm}
	\begin{bmatrix}
		\frac{1}{A_1^2\hspace{-0.02cm}(p)}\mathbf{u}_1\hspace{-0.02cm}(t_k) \\
		\frac{1}{A_2^2\hspace{-0.02cm}(p)}\mathbf{u}_2\hspace{-0.02cm}(t_k) \\
		\vdots \\
		\frac{1}{A_K^2\hspace{-0.02cm}(p)}\mathbf{u}_K\hspace{-0.02cm}(t_k)
	\end{bmatrix}\hspace{-0.06cm}
	\begin{bmatrix}
		\frac{1}{A_1^*(p)A_1(p)}\mathbf{u}_1\hspace{-0.02cm}(t_k) \\
		\frac{1}{A_2^*(p)A_2(p)}\mathbf{u}_2\hspace{-0.02cm}(t_k) \\
		\vdots \\
		\frac{1}{A_K^*(p)A_K(p)}\mathbf{u}_K\hspace{-0.02cm}(t_k)
	\end{bmatrix}^{\hspace{-0.06cm}\top} \hspace{-0.03cm}\right\}\hspace{-0.05cm}, \hspace{-0.05cm}
\end{equation}
and where $\boldsymbol{\mathcal{S}}$ and $\boldsymbol{\mathcal{S}}_*$ are block diagonal matrices formed by $\mathbf{S}(-B_i,A_i)$ and $\mathbf{S}(-B_i^*,A_i^*)$, respectively. Since these block Sylvester matrices are nonsingular, in order for \eqref{inorder} to be generically nonsingular, we must show that $\bm{\Phi}$ is generically nonsingular. Provided the Assumptions \ref{assumption1}, \ref{assumption2} and \ref{assumption4} hold, the generic nonsingularity of such matrix follows from taking $S(q)=1$ and $\mathbf{z}_i(t_k)=\mathbf{u}_i(t_k)$ in Lemma \ref{technicallemma} in the Appendix, which concludes the proof. \hfill $\square$

\textit{Proof of Statement 1, Part (b)}: The proof follows the same lines as the proof of Part (a) of Statement 1 with regards to the perturbation analysis, with $u(t_k)$ replaced by $\tilde{r}(t_k)$ in the expressions. In this case, the noise-free regressor and instrument vectors can be rewritten by considering
\begin{align}
\tilde{\bm{\varphi}}_i^u(t_k,\bm{\beta})&= \mathbf{S}(-B_i^*,A_i^*) S_{uo}^*(q)\frac{1}{A_i(p)A_i^*(p)} \mathbf{r}_{i}(t_k), \\
\hat{\bm{\varphi}}_i(t_k,\bm{\beta})&= \mathbf{S}(-B_i,A_i) S_{uo}(q)\frac{1}{A_i^2(p)}  \mathbf{r}_{i}(t_k),
\end{align}
where $\mathbf{r}_i(t_k)$, $i=1,\dots,K$, have the same structure as \eqref{derivativesu}. Hence, we obtain
\begin{equation}
\overline{\mathbb{E}}\left\{\hat{\bm{\varphi}}(t_k,\bm{\beta})\tilde{\bm{\varphi}}^{r\top}(t_k,\bm{\beta})\right\} = \boldsymbol{\mathcal{S}} \bm{\Phi}_r \boldsymbol{\mathcal{S}}_*^\top, \notag 
\end{equation} 
where $\bm{\Phi}_r$ has the same form as \eqref{phitechnical} in the Appendix, but with $S(q)$ and $\mathbf{z}_i(t_k)$ replaced by $S_{uo}(q)$ and $\mathbf{r}_i(t_k)$, respectively. The generic nonsingularity of this matrix follows from Lemma~\ref{technicallemma}.

\textit{Proof of Statement 2}: As $N\to\infty$, \eqref{nonsingularity} implies that
\begin{equation}
    \label{insert}
    \begin{split}
        \overline{\mathbb{E}}&\left\{ \hat{\bm{\varphi}}(t_k,\bar{\bm{\beta}})\bm{\varphi}^\top(t_k,\bar{\bm{\beta}}) \right\}^{-1} \\
    &\hspace{0.2cm}\times \hspace{-0.03cm} \overline{\mathbb{E}}\hspace{-0.02cm}\left\{ \hat{\bm{\varphi}}(t_k,\bar{\bm{\beta}})\hspace{-0.03cm}\left(\bm{\Upsilon}^\top\hspace{-0.03cm}(t_k,\bar{\bm{\beta}})\hspace{-0.05cm}-\hspace{-0.05cm}\bm{\varphi}^\top(t_k,\bar{\bm{\beta}})\bar{\mathcal{B}} \right) \right\} = \mathbf{0}.
    \end{split}
\end{equation}
Since the matrix inverse above is generically nonsingular by Statement 1, we only need to analyze the second expectation. By following the derivation of the proposed estimator in \eqref{nonsingularity}, we find that 
\begin{align}
\bm{\Upsilon}^\top&(t_k,\bar{\bm{\beta}})-\bm{\varphi}^\top(t_k,\bar{\bm{\beta}})\bar{\mathcal{B}} \notag \\
&= \left(y(t_k)-\sum_{i=1}^K G_i(p,\bar{\bm{\theta}}_i)u(t_k)\right)[1,\dots,1], \notag
    \end{align}
which, when inserted in \eqref{insert}, leads to the condition
\begin{equation}
    \overline{\mathbb{E}}\left\{ \hat{\bm{\varphi}}(t_k,\bar{\bm{\beta}})\left(\sum_{i=1}^K[G_i(p,\bm{\theta}_i^*)- G_i(p,\bar{\bm{\theta}}_i)]u(t_k)\right) \right\} = \mathbf{0}. \notag
\end{equation}
The rest of the proof follows from the derivations made in the proof in Lemma \ref{lemmaconsistency} and is therefore omitted. \hfill $\square$

\section{Simulations}
\label{sec:simulations}
In this section, Monte Carlo experiments are performed to test the asymptotic properties of the proposed estimator, and to compare them with other direct continuous-time identification methods.

\subsection{Open-loop Experiment}
\label{subsec:olexperiment}
\begin{figure}
	\centering{
		\includegraphics[width=0.45\textwidth]{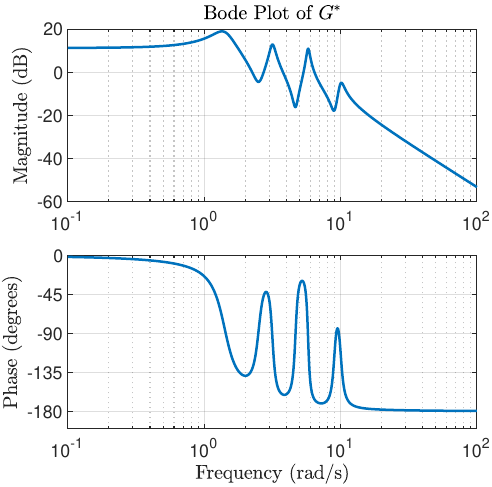}
		\caption{Bode plot of the open-loop continuous-time system under study.}
		\label{fig:bodeol}}
\end{figure}

Consider the following 8th order system
\begin{equation}
    G^*(p) = \sum_{i=1}^4 \frac{b_{i,0}^*}{a_{i,2}^*p^2 + a_{i,1}^*p + 1}, \notag
\end{equation}
where the DC-gains are given by $[b_{1,0}^*, b_{2,0}^*, b_{3,0}^*, b_{4,0}^*] = [3, 0.4, 0.2, 0.05]$, and their associated poles are, respectively, $-0.25\pm 1.39\mathrm{i}$, $-0.15\pm 3.16\mathrm{i}$, $-0.17 \pm 5.77\mathrm{i}$, and  $-0.5 \pm 9.99\mathrm{i}$. The frequency response of this system is shown in Figure~\ref{fig:bodeol}. The input is chosen to be a zero-mean Gaussian noise with unitary variance, and the output is being sampled every $h=0.05$[s]. The noise $v(t_k)$ in~\eqref{output} is given by
\begin{equation}
    v(t_k)= \frac{q+0.5}{q-0.85}e(t_k), \notag 
\end{equation}
where $e(t_k)$ is a zero-mean Gaussian white noise of variance $0.02$, yielding a signal-to-noise ratio of approximately $9$~[dB]. Thirty different sample sizes are considered, ranging logarithmically from $N=2\cdot 10^3$ to $N=10^5$. For each sample size, we conduct $300$ Monte Carlo runs with varying input and noise realizations. The goal is to compare the performance of the proposed estimator with the SRIVC estimator, which is the algorithm behind the \texttt{tfest} command in the MATLAB System Identification Toolbox \cite{ljung2012version}. To this end, we report the mean-square error of the DC gains $b_{i,0}$ related to each submodel. Both estimators initialize at a random system that satisfies $\bm{\beta}_i = \bm{\beta}_i^*(1+\mathcal{U}(-0.05,0.05))$, where $\mathcal{U}(a,b)$ denotes the uniform distribution with lower and upper limits $a$ and $b$, respectively. The maximum number of iterations of the algorithms is set to 100, and the relative error bound that is used as a termination rule for both iterative procedures is set to $10^{-10}$.
\begin{figure}
	\centering{
		\includegraphics[width=0.48\textwidth]{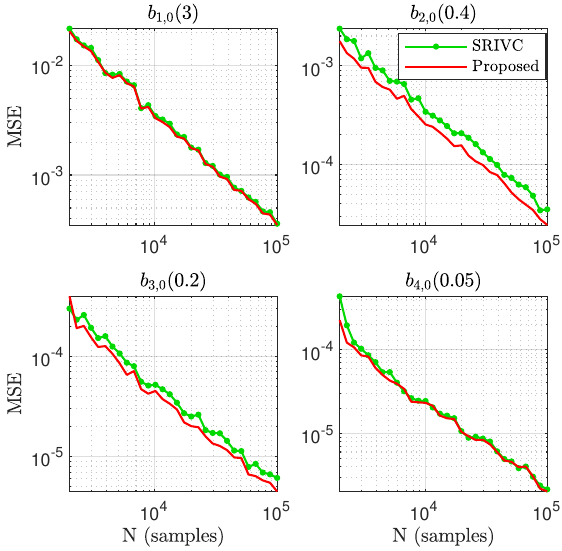}
		\caption{Mean square error of the DC gain estimates with respect to the sample size $N$, open-loop identification. The proposed method delivers parameter estimates with less MSE for $b_{2,0}$ and $b_{3,0}$, while having similar performance to the SRIVC estimator in the other parameters.}
		\label{fig_open_loop}}
\end{figure} 

In Figure \ref{fig_open_loop} we present the MSE of each DC gain estimate as a function of the sample size. Both methods exhibit a linear decay in terms of the MSE of each parameter, which indicates that they are consistent in this scenario. This aligns with Theorem \ref{thmconsistency}. Furthermore, we note a substantial decrease in the MSE for both $b_{2,0}$ and $b_{3,0}$ when comparing the proposed method to the SRIVC method, while remaining competitive with respect to the other parameters. The proposed estimator requires $12$ parameters to describe the system, whereas the standard SRIVC estimator considers $8$ poles and $6$ zeros, leading to a total of $15$ parameters to be estimated. The parsimony of the additive structure results in more accurate estimates.

\subsection{Closed-loop Experiment}
Consider the following 4th order system
\begin{equation}
	\label{gstar}
G^*\hspace{-0.02cm}(p) \hspace{-0.04cm} = \hspace{-0.03cm} \frac{3}{0.25p^2\hspace{-0.03cm}+\hspace{-0.01cm}0.25p\hspace{-0.01cm}+\hspace{-0.02cm}1} + \frac{1}{0.025p^2\hspace{-0.03cm}+0.01p\hspace{-0.01cm}+\hspace{-0.02cm}1},
\end{equation}
which is controlled by a feedback PID controller of the form
\begin{equation}
	C_\textnormal{d}(q) = 0.0115 +\frac{0.00725q}{q-1}+\frac{0.00454(q-1)}{q}, \notag 
\end{equation}
with sampling period $h=0.05[\textnormal{s}]$. The parameters of interest are $\bm{\beta}^* = [0.25,\hspace{0.12cm} 0.25,\hspace{0.12cm} 3, \hspace{0.12cm} 0.01, \hspace{0.12cm} 0.025, \hspace{0.12cm} 1 ]^\top$. The reference is chosen to be a zero-mean Gaussian white noise signal of unitary variance, and the noise $v(t_k)$ is a zero-mean white noise Gaussian of variance $0.01$. Fifty different sample sizes are considered, ranging logarithmically from $N=500$ to $N=123036$, each with $300$ Monte Carlo runs with varying reference and noise realizations. Three estimators are tested: the SRIVC estimator (i.e., the direct closed-loop approach, \cite{forssell1999closed}), the CLSRIVC estimator \cite{gilson2008instrumental}, and the proposed estimator for closed-loop identification, i.e., \eqref{nonsingularity} with instrument vector described by \eqref{instrumentcl}. The estimators are all initialized by the same mechanism detailed in Section \ref{subsec:olexperiment}, and the relative error bound for the termination rule of each algorithm is set to $10^{-7}$. The SRIVC and CLSRIVC estimators consider a model structure consisting of 4 poles and 2 zeros, and the proposed method exploits the parametrization in \eqref{gstar}, which gives $6$ parameters to be computed in total.

\begin{figure}
	\centering{
		\includegraphics[width=0.48\textwidth]{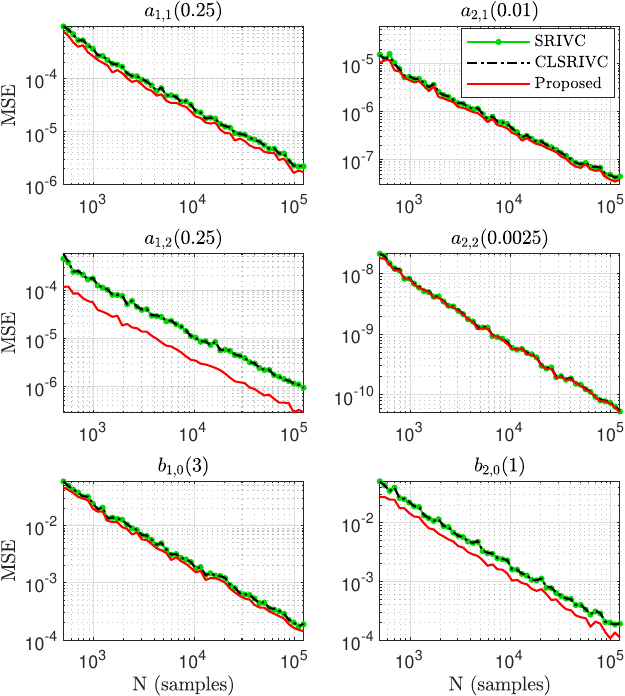}
		\caption{Mean square error of the system parameter estimates with respect to the sample size $N$, closed-loop identification. All estimators give consistent estimates, and the proposed method gives the least mean-square error for every parameter.}
		\label{fig_closed_loop}}
\end{figure}

Our goal is to compare the parameter vectors of the resulting additive form of each estimator. For this, after computing the estimates using the SRIVC and CLSRIVC methods, we factor the resulting transfer function and obtain their associated additive parameter vector estimate. In Figure~\ref{fig_closed_loop}, we have plotted the mean square error of each parameter for all the sample sizes considered in this study. All three estimators are known to be generically consistent in this setup due to Theorems 11 and 15 of \cite{gonzalez2022consistency}, and Theorem \ref{thmconsistency} of this work. However, the proposed method attains lower MSEs compared to the SRIVC and CLSRIVC methods. Similar to Section \ref{subsec:olexperiment}, this improvement is due to a more parsimonious model structure that is imposed: the proposed additive estimator only requires $6$ parameters to exactly describe the system in \eqref{gstar}, as opposed to the unfactored transfer function model descriptions provided by the standard SRIVC and CLSRIVC estimators, which require $7$.

\section{Experimental Validation}
\label{sec:experiments}

\begin{figure}[H]
	\centering{
	\setlength{\fboxsep}{-1pt}%
\setlength{\fboxrule}{1pt}%
    \fbox{\includegraphics[width=243pt]{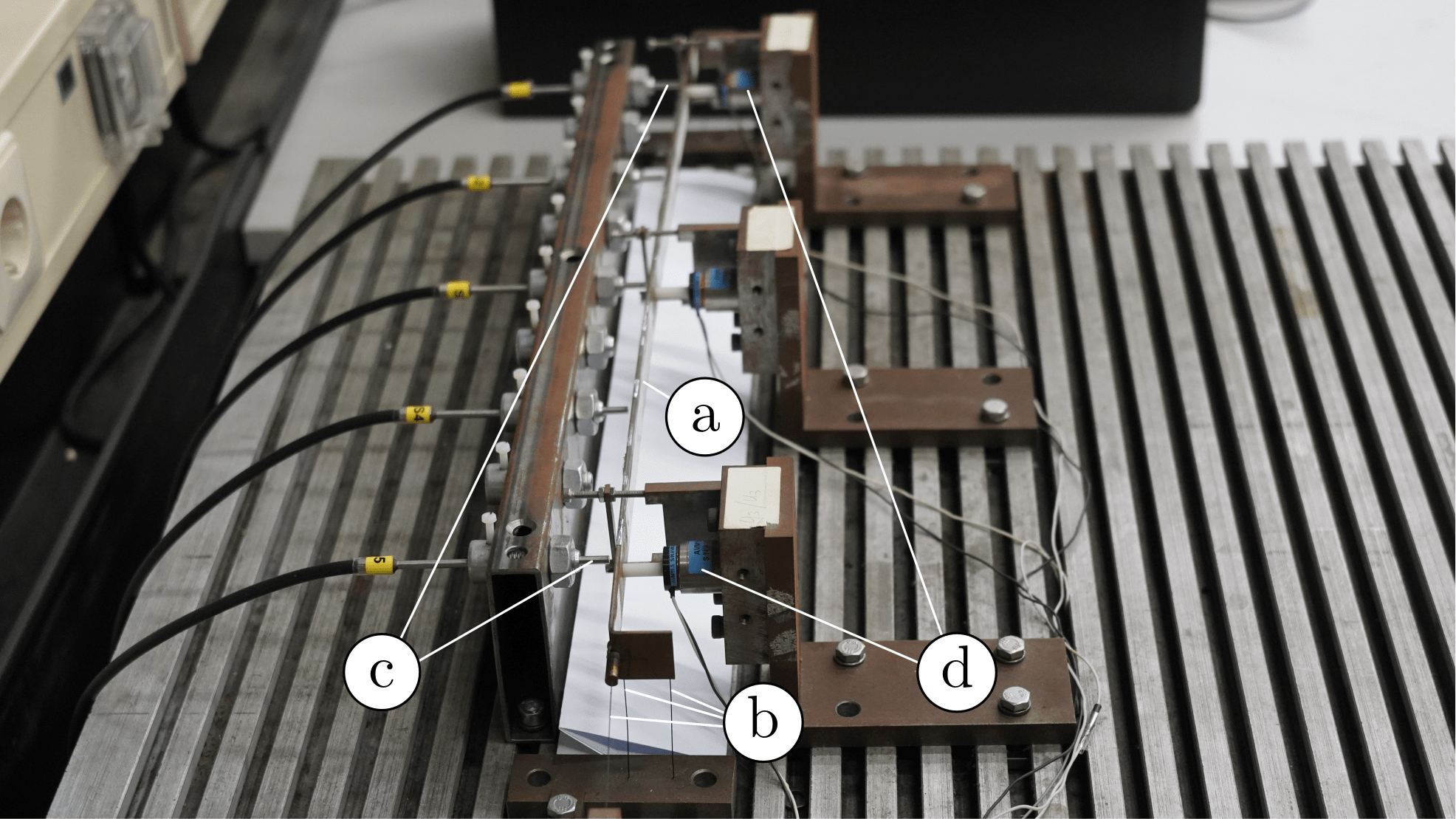}}
\caption{Prototype experimental flexible beam setup. The moving part is indicated by \textcircled{a} and is suspended by wire flexures \textcircled{b}. The deflection is measured with five contactless fiber optic sensors, of which middle sensor is used \textcircled{c} and the setup is actuated with three current-driven voice coils of which the middle actuator used \textcircled{d}.}
		\label{fig_beam}
    }
\end{figure}

In this section, the proposed identification method is validated using experimental data. Consider the setup depicted in Figure \ref{fig_beam}. The system consists of a slender and flexible steel beam of $500 \times 20 \times 2$[mm]. It is equipped with five contactless fiberoptic sensors and three voice-coil actuators and is suspended by wire flexures, leaving one rotational and one translational direction unconstrained. The system is operating at a sampling frequency of $4096$[Hz], and the middle actuator and sensor are used for conducting the experiments. Lightly-damped systems such as the one under study can be described in a modal representation of the form
\begin{equation}
    G^*(p)=\sum_{i=1}^K \frac{b_{i,0}^*}{p^2/\omega_i^2 + 2(\xi_i/\omega_i) p + 1}, \notag 
\end{equation}
where $\xi_i$ and $\omega_i$ represent the relative damping and natural frequencies of the flexible modes, see \cite{gawronski2004advanced} for more details.


\begin{figure}
	\centering{
		\includegraphics[width=0.485\textwidth]{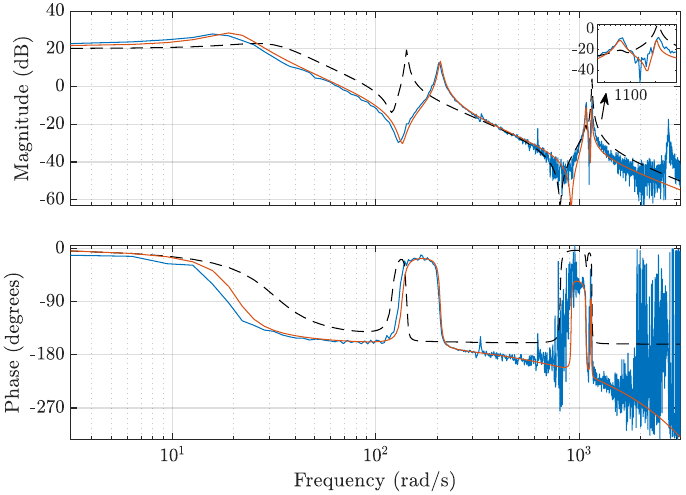}
	\caption{Estimation of the frequency response function of the flexible beam system. A nonparametric estimate \tikzline{MatlabBlue} and the parametric modal estimate with a 4-sample delay \tikzline{MatlabRed} are indicated as well as the initial condition \tikzdashedline{MatlabBlack} of the algorithm. The proposed method converges to a parametric model closely aligned with the first four modes of the system.}
		\label{fig_Beam_estimate_4_modes}}
\end{figure}

An open-loop experiment is conducted with a random-phase multisine input of $2$[s] with a flat spectrum between $0.5$[Hz] to $500$[Hz]. We test the performance of the proposed method for the open-loop scenario, i.e., \eqref{nonsingularity} with instrument vector given by \eqref{instrumentol}. Considering the input and output data $\{u(t_k),y(t_k)\}_{k=1}^{8192}$, the first four modes of the system are estimated (i.e., four second-order systems without zeros), and the result is depicted in Figure \ref{fig_Beam_estimate_4_modes}. The iterations in \eqref{nonsingularity} are initialized with the submodel estimates 
\begin{align}
\bm{\theta}_1^1 &= [ 0.001, 0.025, 10]^\top, &\bm{\theta}_2^1 = [ 5\textnormal{e-5}, 1.5\textnormal{e-4}, 0.2]^\top, \notag \\
\bm{\theta}_3^1 &= [8.5\textnormal{e-7}, 2\textnormal{e-5}, 0.001]^\top,  &\bm{\theta}_4^1 = [7.5\textnormal{e-7}, 5\textnormal{e-6}, 0.01]^\top. \notag 
\end{align}
These parameters describe a model which deviates on purpose from the expected outcome to illustrate the converging nature of the algorithm and its robustness against poor initialization. The delay is estimated to be four samples and the converged estimate of the parametric model has parameters
\begin{align}
\hat{\bm{\theta}}_1 &= [ 0.0024,  0.023, 11.81]^\top, \notag \\ 
\hat{\bm{\theta}}_2 &= [ 2.33\textnormal{e-5}, 1.65\textnormal{e-4}, 0.159]^\top, \notag \\
\hat{\bm{\theta}}_3 &= [8.59\textnormal{e-7}, 1.03\textnormal{e-5}, 3.19\textnormal{e-4}]^\top, \notag \\ 
\hat{\bm{\theta}}_4 &=[7.55\textnormal{e-7}, 3.55\textnormal{e-6}, 1.27\textnormal{e-4}]^{\top}. \notag 
\end{align}
These parameters describe a continuous-time model that is close to the nonparametric estimate obtained using the same data sequence, verifying the validity of the model obtained.
 
\section{Conclusions}
\label{sec:conclusions}
In this paper, we have presented a unified method for identifying continuous-time models in an additive form, applicable to both open and closed-loop settings. This method is derived from the optimality conditions established for both scenarios and extends the properties of well-known refined instrumental variable algorithms to address the identification of systems requiring more flexible model parameterizations. Our open and closed-loop estimators have been rigorously demonstrated to be generically consistent, and extensive simulations were conducted to validate this property. Moreover, our proposed additive identification method has been successfully tested using experimental data. The approach in this paper not only provides accurate models but also offers insights beyond standard unfactored transfer function estimation, as it is capable of directly retrieving the system parameters in its additive form. 

\bibliography{References}

\appendix 	
\section{Technical Lemma}								
\begin{lemma}[Generic Nonsingularity]
\label{technicallemma}
    Consider the $n_i$-th order, asymptotically stable continuous-time transfer functions $1/A_i(p)$, $i=1,\dots,K$, which depend on some parameters $\bm{\nu}_i$, and define $A_i^*(p)$ as the polynomial $A_i(p)$ evaluated at $\bm{\nu}_i=\bm{\nu}_i^*$. The polynomials $A_i^*(p)$, $i=1,\dots,K$, are assumed to be coprime. Furthermore, consider the asymptotically stable discrete-time transfer function $S(q)$, which depends on some parameters $\bm{\xi}$, and define $S^*(q)$ as the transfer function $S(q)$ evaluated at $\bm{\xi}=\bm{\xi}^*$. Assume that $S(q)$ has at most $m_s$ nonminimum phase zeros, and that Assumption \ref{assumption4} holds. If the signal $z(t_k)$ is quasi-stationary and persistently exciting of order no less than $2\sum_{i=1}^K n_i + m_s$, then the following matrix is generically nonsingular with respect to $\{\bm{\nu}_i\}_{i=1}^K$ and $\bm{\xi}$:
    \begin{equation}
    \label{phitechnical}
	\bm{\Phi} \hspace{-0.05cm} = \hspace{-0.05cm}\overline{\mathbb{E}}\hspace{-0.04cm}\left\{\hspace{-0.04cm}
	\begin{bmatrix}
		S\hspace{-0.02cm}(q)\hspace{-0.02cm}\frac{1}{A_1^2\hspace{-0.02cm}(p)}\mathbf{z}_1\hspace{-0.02cm}(t_k) \hspace{-0.05cm} \\
		S\hspace{-0.02cm}(q)\hspace{-0.02cm}\frac{1}{A_2^2\hspace{-0.02cm}(p)}\mathbf{z}_2\hspace{-0.02cm}(t_k) \hspace{-0.05cm} \\
		\vdots \\
		S\hspace{-0.02cm}(\hspace{-0.01cm}q\hspace{-0.01cm})\hspace{-0.02cm}\frac{1}{A_{\hspace{-0.03cm}K}^2\hspace{-0.03cm}(p)}\mathbf{z}_{\hspace{-0.02cm}K}\hspace{-0.03cm}(t_k) \hspace{-0.05cm}
	\end{bmatrix}
 \hspace{-0.07cm}
	\begin{bmatrix}
		\hspace{-0.04cm}S^*\hspace{-0.03cm}(q)\frac{1}{A_1^*(p)A_1(p)}\mathbf{z}_1\hspace{-0.02cm}(t_k) \hspace{-0.04cm} \\
		\hspace{-0.04cm}S^*\hspace{-0.03cm}(q)\frac{1}{A_2^*(p)A_2(p)}\mathbf{z}_2\hspace{-0.02cm}(t_k) \hspace{-0.04cm} \\
		\vdots \\
		\hspace{-0.04cm}S^*\hspace{-0.03cm}(\hspace{-0.01cm}q\hspace{-0.01cm})\frac{1}{A_{\hspace{-0.03cm}K}^*\hspace{-0.03cm}(p)A_{\hspace{-0.03cm}K}\hspace{-0.02cm}(p)}\mathbf{z}_{\hspace{-0.02cm}K}\hspace{-0.03cm}(\hspace{-0.01cm}t_k\hspace{-0.01cm}) \hspace{-0.04cm}
	\end{bmatrix}^{\hspace{-0.06cm}\top}\hspace{-0.03cm} \right\}\hspace{-0.04cm}, 
\end{equation}
where $\mathbf{z}_i(t_k) = [p^{n_i+m_i}, p^{n_i+m_i-1},\dots, 1 ]^\top z(t_k)$, with $m_i<n_i$ for all but at most one $i$, at which $m_i=n_i$.
\end{lemma}

\textit{Proof}:
\label{appendixlemmaphi}
We need to show that the two statements needed for applying the genericity result in Lemma A2.3 of \cite{soderstrom1983instrumental} are true, which are that
\begin{enumerate}
	\item all the entries of $\bm{\Phi}$ are real analytic functions of the coefficients of  $A_i(p)$ and $S(q)$, and
	\item there exist $\{\bm{\nu}_i\}_{i=1}^K$ and $\bm{\xi}$ vectors that lead to $\bm{\Phi}$ being nonsingular.
\end{enumerate}
The first statement follows from applying the same logic as in Lemma 9 of \cite{pan2020consistency} and is therefore omitted. As for the second statement, we will show that setting $S(q)=S^*(q)$ and $T_i(p) = T_i^*(p)$ for all $i=1,2,\dots,K$, leads to a nonsingular matrix $\bm{\Phi}$. To this end, we define
\begin{equation}
	\bm{\Phi}^* \hspace{-0.09cm}= \hspace{-0.06cm}\overline{\mathbb{E}}\hspace{-0.06cm}\left\{\hspace{-0.09cm}
	\begin{bmatrix}
		S^*\hspace{-0.03cm}(q)\frac{1}{[A_1^*(p)]^2}\mathbf{z}_1\hspace{-0.02cm}(\hspace{-0.01cm}t_k\hspace{-0.01cm}) \\
		S^*\hspace{-0.03cm}(q)\frac{1}{[A_2^*(p)]^2}\mathbf{z}_2\hspace{-0.02cm}(\hspace{-0.01cm}t_k\hspace{-0.01cm}) \\
		\vdots \\
		S^*\hspace{-0.03cm}(q)\frac{1}{[A_K^*\hspace{-0.02cm}(p)]^2}\mathbf{z}_K\hspace{-0.02cm}(\hspace{-0.01cm}t_k\hspace{-0.01cm})
	\end{bmatrix}\hspace{-0.11cm}
	\begin{bmatrix}
		S^*\hspace{-0.03cm}(q)\frac{1}{[A_1^*(p)]^2}\mathbf{z}_1\hspace{-0.02cm}(\hspace{-0.01cm}t_k\hspace{-0.01cm}) \\
		S^*\hspace{-0.03cm}(q)\frac{1}{[A_2^*(p)]^2}\mathbf{z}_2\hspace{-0.02cm}(\hspace{-0.01cm}t_k\hspace{-0.01cm}) \\
		\vdots \\
		S^*\hspace{-0.03cm}(q)\frac{1}{[A_K^*\hspace{-0.02cm}(p)]^2}\mathbf{z}_{\hspace{-0.02cm}K}\hspace{-0.02cm}(\hspace{-0.01cm}t_k\hspace{-0.01cm})
	\end{bmatrix}^{\hspace{-0.08cm}\top} \hspace{-0.04cm}\right\}\hspace{-0.06cm}. \notag
\end{equation}
Take $\mathbf{x}\in \mathbb{R}^{K+\sum_{i=1}^K (n_i+m_i)}$. The following inequality is direct: 
\begin{equation}
	\label{alsowrite}
	\mathbf{x}^\top \bm{\Phi}^* \mathbf{x} = \overline{\mathbb{E}}\left\{\left(\frac{B_\mathbf{x}(p)}{\prod_{i=1}^{K} [A_i^*(p)]^2}\tilde{z}(t_k)\right)^2\right\} \geq 0,
\end{equation}
where $\tilde{z}(t_k) = S^*(q)z(t_k)$ is a persistently exciting signal of order no less than $2\sum_{i=1}^K n_i$, and $B_\mathbf{x}(p)$ is a polynomial of degree at most $2\sum_{i=1}^{K}n_i-\max_{k} (n_k-m_k)$ of the form
\begin{equation}
	B_\mathbf{x}(p) = \sum_{i=1}^K Q_i(p)\prod_{\substack{j = 1,\dots,K,\\ j \neq i}} [A_j^*(p)]^2, \notag 
\end{equation}
with $Q_i$ ($i=1,\dots,K$) being an arbitrary polynomial of degree $n_i+m_i$, described by the entries of the vector $\mathbf{x}$. By following the same arguments as in Eqs. (38-40) of \cite{pan2020consistency} (which require the persistence of excitation and sampling frequency assumptions), we find that $\mathbf{x}^\top \bm{\Phi}^* \mathbf{x}=0$ implies that $B_\mathbf{x}(p)\equiv 0$. 

Now our goal is to show that $B_\mathbf{x}(p)\equiv 0$ implies that $Q_i(p)\equiv 0$ for all $i=1,\dots,K$, which means that $\mathbf{x}=\mathbf{0}$. First, it is clear that at least two $Q$ polynomials must be nonzero in case the statement were to be disproven. To tackle this case, we argue by contradiction. Suppose that there are at least two polynomials $Q_k, Q_l$ that are nonzero, and without loss of generality\footnote{At most one polynomial $Q_i$ will have degree $2n_i$, due to $n_i=m_i$ occurring for at most one $i\in \{1,\dots,K\}$.} assume that $\text{deg}(Q_k)<2n_k$. Since $B_\mathbf{x}(p)\equiv 0$, the following relations between polynomials hold true:
\begin{align}
	\label{lefthandside}
	Q_{k}(p)&\prod_{\substack{j = 1,\dots,K,\\ j \neq k}} [A_j^*(p)]^2 \\
	&= -[A_k^*(p)]^2\sum_{\substack{i = 1,\dots,K,\\ i \neq k}} Q_i(p)\prod_{\substack{j = 1,\dots,K,\\ j \neq i,k}} [A_j^*(p)]^2 \notag \\
	&\neq 0. \notag
\end{align}
Thus, the polynomial in \eqref{lefthandside} must have the same zeros as $[A_k^*(p)]^2$, counting their multiplicity. These zeros cannot be accounted for in $\prod_{\substack{j = 1,\dots,K,\\ j \neq k}} [A_j^*(p)]^2$ since the $A_j^*$ polynomials are jointly coprime, and neither in $Q_k(p)$, since $\text{deg}(Q_k)<\text{deg}([A_k^*(p)]^2)=2n_k$. We conclude that such $Q_k,Q_l$ nonzero polynomials that satisfy the equality in \eqref{lefthandside} cannot exist. Hence, by contradiction, we have found that all $Q$ polynomials must be zero for $B_\mathbf{z}(p)\equiv 0$ to be true, and therefore,  $\bm{\Phi}^*$ is nonsingular.

Since both statements 1) and 2) are true, the result in Lemma \ref{technicallemma} follows from Lemma A2.3 of \cite{soderstrom1983instrumental}. \hfill $\square$

\end{document}